\RequirePackage{fix-cm}

\documentclass[smallextended]{svjour3}       
\smartqed  
\usepackage{times,latexsym,euscript,amstext,amssymb,amsbsy,amsmath,amsfonts,bm}
\usepackage{graphicx}

\def\s#1{\scriptscriptstyle{#1}}

\def\noeq#1{(\ref{#1})}
\def\1eq#1{Eq.~(\ref{#1})}

\def\2eqs#1#2{Eqs.~(\ref{#1}) and~(\ref{#2})}
\def\3eqs#1#2#3{Eqs.~(\ref{#1}),~(\ref{#2}) and~(\ref{#3})}

\def\fig#1{Fig.~\ref{#1}}

\def\Cgh{\widetilde{C}_{\mathrm{gh}}}
\def\Cgl{\widetilde{C}_{\mathrm{gl}}}
\def\g{\widetilde\Gamma^{ \bf np}}

\def\hh{m^2_{\mathrm{gl}}}
\def\diff{{\rm d}}
\def\aBSE{\alpha_s^{\s{\mathrm{BSE}}}}
\def\aDSE{\alpha_s^{\s{\mathrm{DSE}}}}

\def\fQ{f_{\mathrm{gl}}}

\begin{document}

\title{Emergent Hadron Mass in Strong Dynamics}
\author{Daniele Binosi}
\institute{Daniele Binosi \at
              ECT*, Strada delle Tabarelle 256, I-28213, Villazzano, Italy \\
              Tel.: +0039 0461 314 738\\
              \email{binosi@ectstar.eu}           
}

\date{Received: date / Accepted: date}

\maketitle

\begin{abstract}
Emergent Hadron Mass (EHM) is a mechanism capable of explaining both the unnaturally small (pion) and large (proton) masses of hadrons and if not the origin of confinement, then intimately connected with it. Even though the modern formulation of EHM has only recently been completed, it is rapidly becoming a fundamental focus for modern and future experimental efforts aimed at understanding the strong force within the Standard Model. Using Dyson-Schwinger equations for the gluon, ghost and quark propagators, I will introduce the main EHM concepts and illustrate how far this framework can take us in understanding strong interaction phenomeno- logy.

\end{abstract}

\section{Introduction}

Quantum ChromoDynamics (QCD) is arguably the most fascinating and challenging part of the Standard Model of particle physics. Its lowest mass bound state, the proton, is a composite object comprised by two up and one down quarks bound together through the exchange of gluons. The proton has never decayed in the nearly 14-billion years since the Big Bang; and this extraordinarily long lifetime is basic to the existence of all known matter: you wouldn't be reading this (nor I would have written it) if that were not the case. 

QCD is also the only self-consistent theory that is known to us: so far, never was there the need to add or change anything in it through an enormous energy range spanning 10 orders of magnitude; and, due to its asymptotically free nature~\cite{Gross:1973id,Politzer:1973fx}, it is unlikely that it will break down at any energy scale at all. Finally, there is no intrinsic parameter to fine tune, only one observable to be measured to set a single scale (what is known as $\Lambda_{\mathrm{\s{QCD}}}$). QCD is thus the only known example of a theory rather than an {\it effective} theory.

The flip side of the coin is that asymptotic freedom implies that the theory is innately nonperturbative at short distances, which in turn means that it is extremely difficult to describe QCD's dynamics at the fm scale. Indeed, QCD's colour-charged degrees of freedom, the massless gluons and the light up and down quarks, have never been observed in isolation: they are confined in colour-neutral bound-states, the hadrons, of which the (light) pion and the (heavy) proton are primary examples. It should be noticed that without a mass-scale confinement would not be possible: colour-singlet combinations of quarks would still be there, but the participating particles would need not be close together, since in a scale invariant theory all lengths are equivalent. Accordingly, the question ``how does confinement appear in QCD'' is inextricably connected to the question ``how does mass emerge in strong dynamics''. 

You might believe that the Higgs boson, found experimentally in 2012~\cite{ATLAS:2012yve,CMS:2012qbp}, is the answer; but you would be wrong. The Higgs produces a tiny mass for the $u$ and $d$ quarks, roughly $m_u\approx m_d/2\approx0.0022$ GeV. The (positive) pion, the lightest of hadrons, which contains one $u$ and one $\bar d$ (valence) quarks, has a mass of 0.140 GeV which is 20 times bigger than the sum of its components; similarly, the proton mass clocks at a staggering 0.938 GeV which is  140 times bigger than the sum of the masses of its two $u$ and one $d$ (valence) quarks. Explaining how exactly this mass generation mechanism works would be nothing short than understanding how $\sim$98\% of the mass in the visible Universe came into being $10^{-6}$ seconds after the Big-Bang. This is the time when the phenomenon of Emergent Hadron Mass (EHM)~\cite{Roberts:2020hiw} starts. The massless gluon and quark elementary particles appearing in QCD's gauge-invariant Lagrangian (see below) are first turned by strong interactions into complex quasiparticles, characterized by a dynamically generated momentum-dependent mass-function whose value is large at IR momenta (roughly one-half and one-third of the $\sim$1 GeV proton mass). Next, these quasiparticles form massive bound-states, with EHM imprinted in their associated properties and empirical observables. 

In the following I will describe a framework in which EHM can be rigorously tackled and understood, and describe some of its most immediate consequences.  

\paragraph{Quantum ChromoDynamics.} The wonderfully simple Lagrangian density of a SU($N$) Yang-Mills theory can be  written as the sum of three terms:
\begin{align}
	{\cal L}={\cal L}_{\mathrm{I}}+{\cal L}_{\mathrm{GF}}+{\cal L}_{\mathrm{FPG}}.
	\label{lagden}
\end{align}
${\mathcal L}_{\mathrm I}$ represents the gauge invariant SU(3) density
\begin{equation}
{\mathcal L}_{\mathrm I} = -\frac14 F_a^{\mu\nu}F^a_{\mu\nu}+\bar{q}^i_\mathrm{f}
\left(i\gamma^\mu{\mathcal D}_\mu-m\right)_{ij}q^j_\mathrm{f} ,
\label{Linv}
\end{equation}
where $a=1,\dots,N^2-1$ (respectively $i,j=1,\dots,N$) is the color index for the adjoint (respectively fundamental) representation, while ``f'' is the flavor index. 
The field strength is
\begin{equation}
F^a_{\mu\nu}=\partial_\mu A^a_\nu-\partial_\nu A^a_\mu+gf^{abc}A^b_\mu A^c_\nu,
\end{equation}
with $f^{abc}$ the totally antisymmetric structure constants appearing in the commutation relations satisfied by the SU$(N)$ generators $t^a$, namely
\begin{equation}
[t^a,t^b]=if^{abc}t^c.
\label{SU(N)_gen_comm_rel}
\end{equation}
Finally, the covariant derivative is defined according to
\begin{equation}
({\mathcal D}_\mu)_{ij}=\partial_\mu (\mathbb{I})_{ij}-ig A^a_\mu (t^a)_{ij},
\end{equation}
with $g$ the (strong) coupling constant. 

The last two terms in~\1eq{lagden} represent the gauge-fixing  and Faddev-Popov ghost terms, respectively. The most general way of writing them is by introducing a gauge-fixing function ${\cal F}^a$ and coupling it to a set of Lagrange multipliers $b^a$ (the so-called Nakanishi-Lautrup multipliers~\cite{Nakanishi:1966cq,Lautrup:1966cq}). One then has
\begin{align}
	{\cal L}_{\mathrm{GF}}+{\cal L}_{\mathrm{FPG}}=s\left[\overline c^a{\cal F}^a-\frac\xi2\overline c^a b^a\right],
	\label{GF+FPG}
\end{align}
where: $\overline c^a$ (and, respectively, $c^a$ appearing below) are the anti-ghost (ghost) fields; $\xi$ is a (non-negative) gauge-fixing parameter; and, finally, $s$ is the Becchi-Rouet-Stora-Tyutin  (BRST) operator~\cite{Becchi:1974md,Becchi:1975nq,Tyutin:1975qk}, which acts on the various fields according to
\begin{align}
	sA^a_\mu&=\partial_\mu c^a+gf^{abc}A^b_\mu c^c;&
	sc^a&=-\frac12f^{abc}c^bc^c;&
	s\bar c^a&=b^a;&
	sb^a&=0,\nonumber \\
	sq^i_\mathrm{f}&=ig c^a(t^a)_{ij}q^j_\mathrm{f};&
	s\bar q^i_\mathrm{f}&=-ig c^a\bar q^j_\mathrm{f} (t^a)_{ji},&
	\label{BRST}
\end{align}
Notice that the $b^a$ fields have no dynamical content and can be eliminated through their (trivial) equations of motion. In the ubiquitous renormalizable $\xi$ gauges ($R_\xi$ gauges for short), one chooses~\cite{Fujikawa:1972fe}
\begin{align}
	{\cal F}^a=\partial^\mu A^a_\mu,
	\label{lincov}
\end{align}
so that, going on-shell with the $b$ field, we obtain the familiar result\footnote{Observe that on-shell the $b$ field is such that $\xi b^a={\cal F}^a$; additionally, in the adjoint representation one has $(t^a)_{bc}=-if^{abc}$.}
\begin{align}
	{\cal L}_{\mathrm{GF}}+{\cal L}_{\mathrm{FPG}}=-\bar c^a\partial^\mu{\cal D}^{ab}_\mu c^b+\frac1{2\xi}(\partial^\mu A^a_\mu)^2.
\end{align}
Feynman rules derived from the Lagrangian density~\noeq{lagden} are given in Appendix B of~\cite{Binosi:2009qm}.

\begin{figure}
\centering
\includegraphics[scale=0.65]{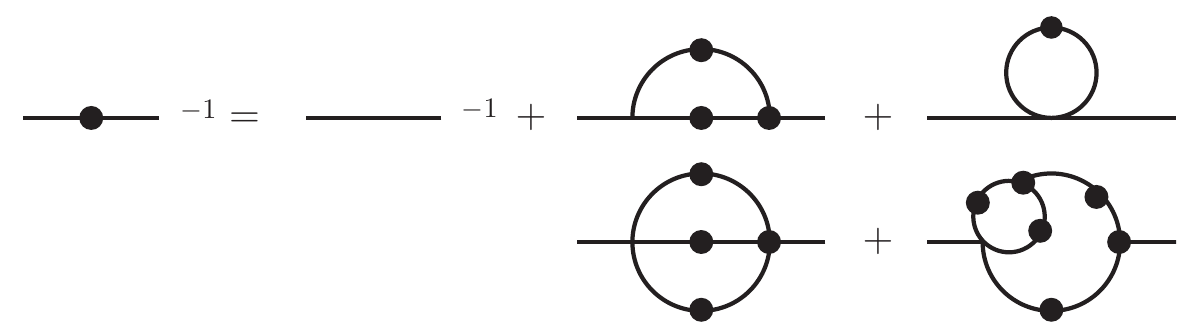}
\caption{\label{fig:DSE}The topologies contributing to the QCD 2-point functions' DSE. Black blobs correspond to fully dressed functions. Replacement of the lines using the allowed Feynman rules constructed from the Lagrangian density~\noeq{lagden} allows the derivation of the gluon, ghost and quark propagator DSE. In particular, in the gluon case all topologies contribute, whereas for the ghost and quark case only the first non-trivial topology of the first line is active.}	
\end{figure}

\paragraph{Dyson-Schwinger equations.} From the generating functional associated with the action $S=\int\!\mathrm{d}^dx{\cal L}$ one can then derive (see, {\it e.g.},~\cite{Roberts:1994dr} and references therein) the Dyson-Schwinger equations (DSEs) of the theory, which are simply the Euler-Lagran- ge equations of motion of the quantum fields. These equations are valid in both the ultraviolet (UV) and infrared (IR) regimes, for weak and strong coupling values, and independently of the masses of the particles in and out the loops; and, since they constitute an infinite tower of coupled integral equations relating the different $n$-point functions, solving these equations allows to reconstruct the entire generating functional, thereby solving completely the theory. As an example, in \fig{fig:DSE} I show the topologies appearing in the DSE corresponding to QCD's 2-point sector (the propagators); the corresponding Feynman rules can be then used to explicitly construct from this skeleton expansion the three DSEs corresponding to the gluon, ghost and quark propagator. Notice that already at the 2-point level (the lowest possible in QCD) one sees the anticipated appearance of higher order functions, specifically the 3- and 4-point functions. 

Thus any study based on DSEs unavoidably involves the specification of a truncation scheme, {\it i.e.}, the specification of the maximum number of legs $n$ which will be treated self-consistently through their own DSE, usually employing an Ansatz for all the remaining functions. A lot of progress has been made in the last 20 years in developing truncation schemes capable of maintaining intact certain desirable properties of the theory including: various global and local symmetries; known perturbative behaviour; (multiplicative) renormalizability; analyticity, etc. When coupled with the additional insights into the IR behavior of 2- and 3-point functions provided by lattice simulations in the same time-span, this has allowed DSEs studies to produce {\it model-independent statements about QCD}. I will give an illustration of many such statements in the rest of these lectures, studying mainly 2- and 3-point functions.

\newpage

\section{Gauge 2-point sector}

\subsection{Gluons}

\begin{figure}[!t]
	\centering
	\includegraphics[scale=0.45]{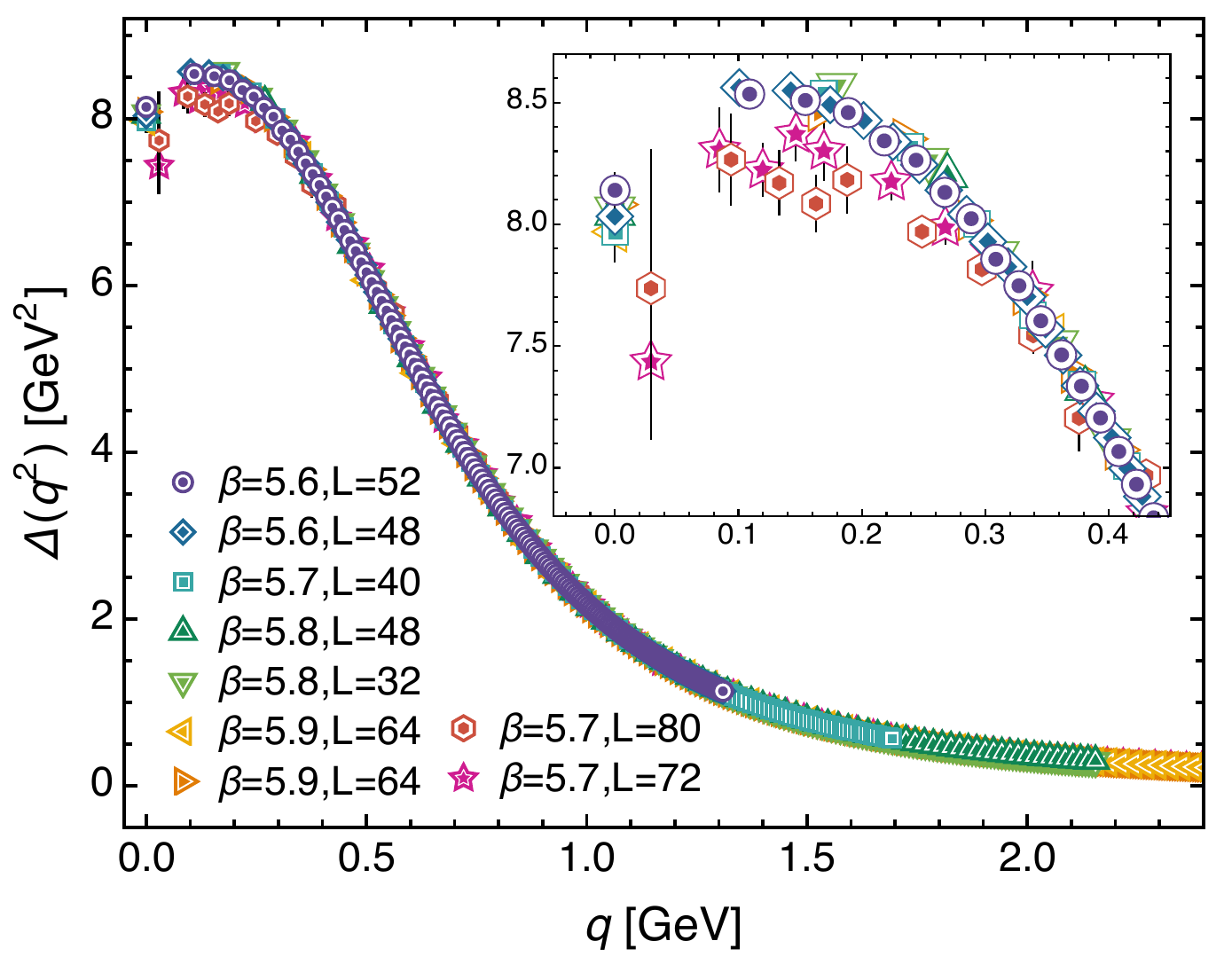}
	\caption{\label{fig:Delta-latt}Compilation of the latest SU$(3)$ quenched lattice results for the gluon propagator in the Landau gauge corrected for discretization artefacts~\cite{Aguilar:2021okw}. The inset shows the IR region; notice the saturation to a finite non-vanishing value and the presence of a maximum, which, as we will see, is a consequence of the gluon-ghost dynamics. Data are renormalized at $\mu=4.3$ GeV; given a lattice propagator function $I$ renormalized at a point $\zeta$, renormalization at a different point $\mu$ can be achieved through the formula $I(q^2;\mu^2)=I(q^2;\zeta^2)/[\mu^2I(\mu^2;\zeta^2)]$.}
\end{figure}

The first DSE I will analyze is also the most difficult one, {\it i.e.}, the one corresponding to the gluon propagator. The reason why this DSE is so interesting is because it was  suggested forty years ago that a Schwinger-like mechanism, by which a gauge boson may acquire a mass provided that its vacuum polarization function develops a pole at zero momentum transfer~\cite{Schwinger:1962tn,Schwinger:1962tp}, might be active in QCD \cite{Cornwall:1981zr}. This would in turn mean that the gluon propagator would saturate at a non-vanishing value in the deep IR region, thus signalling that gluons acquire a dynamical momentum dependent mass due to their self-interactions~\cite{Cornwall:1981zr,Aguilar:2008xm}. This prediction has been spectacularly confirmed by large-volume lattice simulations (\fig{fig:Delta-latt}) for different gauge groups [SU(3) or SU(2)],  (covariant) gauges, and the absence/presence of dynamical quarks~\cite{Cucchieri:2007md,Cucchieri:2007rg,Bowman:2007du,Bogolubsky:2007ud,Bogolubsky:2009dc,Oliveira:2009eh,Cucchieri:2009zt,Cucchieri:2010xr,Ayala:2012pb,Binosi:2016xxu,Bicudo:2015rma}. 

In the following\footnote{For other approaches to solving the DSEs in general and understanding the IR saturation of the gluon propagator in particular, see~\cite{Dudal:2008sp,Tissier:2011ey,Cyrol:2016tym}. A detailed summary is also provided in~\cite{Huber:2018ned}.}, using gluon's DSE emerging from the combination of the Pinch Technique (PT)~\cite{Binosi:2009qm,Cornwall:1981zr,Cornwall:1989gv,Binosi:2002ft,Binosi:2003rr,Binosi:2004qe} with the Background Field Method (BFM)~\cite{Abbott:1980hw,Abbott:1981ke} (simply referred to as ``PT-BFM''~\cite{Aguilar:2006gr,Binosi:2007pi,Binosi:2008qk}), I will argue that gauge sector dynamics transforms the massless gluon partons in~\1eq{Linv} into complex quasiparticles, characterized by a momentum-dependent mass-function whose value is large at IR momenta. As we will see, the Schwinger-like $1/q^2$ pole in the self-energy can only emerge in QCD because a long-range (massless) longitudinally-coupled coloured correlation is dynamically generated in QCD's three-gluon vertex (and possibly ghost-gluon and four-gluon ones)~\cite{Jackiw:1973tr,Jackiw:1973ha,Cornwall:1973ts,Eichten:1974et,Poggio:1974qs}.

\paragraph{One-loop Pinch Technique gluon self-energy\label{1lPT}.}

To begin with, let me consider the gluon self-energy at one-loop; neglecting quark's contribution, it reads\footnote{I use for convenience the Feynman gauge, $\xi=1$; a similar construction to the one following can be carried out in an arbitrary $R_\xi$ gauge using the so-called generalized Pinch Technique algorithm~\cite{Pilaftsis:1996fh}.}
\begin{align}
	\Pi^{(1)}_{\alpha\beta}(q;\mu) &=\Pi^{\rm{gl},(1)}_{\alpha\beta}(q;\mu)+\Pi^{\rm{gh},(1)}_{\alpha\beta}(q;\mu)\nonumber \\
	&=\frac12g^2C_{A}\int_k \frac{\Gamma^{(0)}_{\alpha\mu\nu}(q,k,-k-q) 
\Gamma_{\beta}^{(0)\mu\nu}(-q,k+q,-k)}{k^2 (k+q)^2}\nonumber \\
&\mathrel{\phantom{=}}-\frac12g^2C_{A}\int_k\!\frac{
k_{\alpha}(k+q)_{\beta}+k_{\beta}(k+q)_{\alpha}}{k^2 (k+q)^2},	
\label{1loopglseconv}
\end{align}
where: $C_{A}$ is the Casimir eigenvalue of the adjoint representation of the SU$(N)$ gauge group ($C_{A}=N$); I have introduced the short-hand dimensional regularization notation
\begin{equation}
	\int_{k}\equiv\frac{\mu^{2\varepsilon}}{(2\pi)^{d}}\int\!{\mathrm d}^d k,
\end{equation}
with $d=4-\epsilon$ the dimension of space-time and $\mu$ the 't Hooft mass-scale which guarantees that the coupling constant is dimensionless in $d$ dimensions; $\Gamma_{\alpha\mu\nu}$ is the three-gluon vertex 
\begin{subequations}
\begin{align}
	i\Gamma^{amn}_{\alpha\mu\nu}(q,k_1,k_2)&=g f^{amn}
\Gamma_{\alpha \mu \nu}(q,k_1,k_2),  \\
\Gamma^{(0)}_{\alpha \mu \nu}(q,k_1,k_2)& = g_{\mu\nu}(k_1-k_2)_{\alpha}+g_{\alpha\nu} (k_2-q)_{\mu}+
g_{\alpha\mu}(q-k_1)_{\nu},
\label{tgv_ch2}
\end{align}
\label{tgv-glob}
\end{subequations}
\hspace{-0.18cm} with all momenta incoming ($q+k_1+k_2 = 0$); and, for later convenience, the ghost contribution has been symmetrized. Now\footnote{To avoid notational clutter I will omit the explicit dependence of a Green function on the renormalization scale $\mu$. The dependence will be reinstated when needed for clarity, {\it e.g.}, when dealing explicitly with renormalization and renormalization group invariance issues.}, it is easy to show that the full gluon self-energy, given at one-loop by the sum of the gluon and ghost diagrams, is transverse, {\it i.e.},
\begin{align}
	&q^\beta\Pi^{(1)}_{\alpha\beta}(q)=0& &\Rightarrow& 
	&\Pi^{(1)}_{\alpha\beta}(q)=P_{\alpha\beta}(q)\Pi^{(1)}(q^2);& P_{\alpha\beta}(q)=g_{\alpha\beta}-q_\alpha q_\beta/q^2,
\end{align}
which is in fact an all-order identity enforced by gauge symmetry. However, this property is not satisfied by the gluon and ghost contributions in isolation, that is
\begin{align}
	&q^\beta\Pi^{\mathrm{gl},(1)}_{\alpha\beta}(q)\neq0;&\qquad  &q^\beta\Pi^{\mathrm{gh},(1)}_{\alpha\beta}(q)\neq0.
	\label{ortrans}
\end{align}

I now split the tree-level three-gluon vertex into two parts, $\Gamma^{(0)}_{\alpha \mu \nu}=\widetilde{\Gamma}^{(0)}_{\alpha \mu \nu}+\Gamma_{\alpha \mu \nu}^{{\rm P}}$, with 
\begin{subequations}
\begin{align}
\widetilde{\Gamma}^{(0)}_{\alpha \mu \nu}(q,k_1,k_2) &=
(k_1-k_2)_{\alpha} g_{\mu\nu} + 2q_{\nu}g_{\alpha\mu} 
- 2q_{\mu}g_{\alpha\nu}, \label{GF}\\
\Gamma_{\alpha \mu \nu}^{{\rm P}}(q,k_1,k_2) &= 
 k_{2\nu} g_{\alpha\mu} - k_{1\mu}g_{\alpha\nu}, 
\label{GP}
\end{align}
\end{subequations}
and proceed to carry out the following rearrangement of the two tree-level three-gluon vertices appearing in $\Pi^{\mathrm{gl},(1)}_{\alpha\beta}$:
\begin{align}
	\Gamma^{(0)}_{\alpha \mu \nu}\Gamma^{(0)\mu \nu}_{\beta}  &=
\widetilde{\Gamma}^{(0)}_{\alpha \mu \nu}\widetilde{\Gamma}^{(0)\mu \nu}_{\beta}
+\Gamma^{{\rm P}}_{\alpha \mu \nu}\Gamma^{(0)\mu \nu}_{\beta}
+\Gamma^{(0)}_{\alpha \mu \nu}\Gamma^{{\rm P}\,\mu \nu}_{\beta}
-\Gamma^{{\rm P}}_{\alpha \mu \nu} \Gamma^{{\rm P}\,\mu\nu}_{\beta}.
\label{INPTDEC1}
\end{align}
It is then immediate to prove that, with the momenta routing of~\1eq{1loopglseconv},
\begin{subequations}
\begin{align}
\Gamma^{{\rm P}}_{\alpha \mu \nu}\Gamma_{\beta}^{(0)\mu \nu}+
\Gamma^{(0)}_{\alpha\mu\nu}\Gamma^{{\rm P}\,\mu\nu}_{\beta}
&= - 4 q^2 P_{\alpha\beta}(q) - 
2 k_{\alpha} k_{\beta} - 2 (k+q)_{\alpha}(k+q)_{\beta}, \label{GPG+GGP}\\
\Gamma^{{\rm P}}_{\alpha \mu \nu}\Gamma^{{\rm P}\,\mu\nu}_{\beta}  
&= 2 k_{\alpha}k_{\beta}+ (k_{\alpha}q_{\beta}+q_{\alpha}k_{\beta}),
\label{GPGP}
\end{align}
\end{subequations}
where $P_{\alpha\nu}(q)=g_{\alpha\beta}-q_\alpha q_\beta/q^2$ is the transverse projector, and several terms have been set to zero in view of the use of dimensional regularization result
\begin{align}
	\int_k \frac{1}{k^{2}} = 0.
\end{align}
Thus I obtain\footnote{This is a first example of what will be later called a Background-Quantum identity.} 
\begin{align}
	\Pi^{(1)}_{\alpha\beta}(q) &= \frac12g^2C_{A} 
\int_k\!\frac{\widetilde{\Gamma}^{(0)}_{\alpha \mu \nu}(q,k,-k-q)\widetilde{\Gamma}^{(0)\mu \nu}_{\beta}(-q,k+q,-k)}{k^2 (k+q)^2}\nonumber \\
&\mathrel{\phantom{=}}-g^2C_{A}\int_k\!\frac{\widetilde{\Gamma}^{(0)}_{\alpha}(q,k,-k-q)\widetilde{\Gamma}^{(0)}_{\beta}(-q,k+q,-k)}{k^2 (k+q)^2}\nonumber\\
&\mathrel{\phantom{=}}-2g^2C_{A}q^2 P_{\alpha\beta}(q) \int_k\!\frac{1}{k^2 (k+q)^2}\nonumber \\
&=\widehat{\Pi}^{\mathrm{gl},(1)}_{\alpha\beta}(q)+\widehat{\Pi}^{\mathrm{gh},(1)}_{\alpha\beta}(q)-2\Pi^{P,(1)}_{\alpha\beta}(q)=\widehat{\Pi}^{(1)}_{\alpha\beta}(q)-2\Pi^{P,(1)}_{\alpha\beta}(q),
\label{propexp}
\end{align}
where I have defined the modified gluon-ghost vertex 
\begin{align}
	\widetilde{\Gamma}^{(0)}_{\alpha}(q,k_1,k_2)=(k_2-k_1)_\alpha.
	\label{Bcbc}
\end{align}
The gluon self-energy has been thus decomposed into three pieces: a modified gluon and ghost contribution which are obtained from the original diagrams by defining somewhat different Feynman rules for the three-gluon and gluon-ghost vertices; and a `pinch' term. As for the pinch term $\Pi^{P,(1)}_{\alpha\beta}$, the (intrinsic) PT prescription indicates~\cite{Cornwall:1981zr,Cornwall:1989gv} to discard all pieces proportional to the transverse combination $q^2 P_{\alpha\beta}$ generated from the three-gluon vertex decomposition $\Gamma^{(0)}_{\alpha \mu \nu}=\widetilde{\Gamma}^{(0)}_{\alpha \mu \nu}+\Gamma_{\alpha \mu \nu}^{{\rm P}}$. This is because such pieces would cancel with similar propagator-like contributions coming from different diagrams when constructing the PT gluon self-energy by embedding it in a one-loop $S$-matrix scattering process like quark+quark elastic scattering (this goes under the name of $S$-matrix PT~\cite{Cornwall:1981zr,Cornwall:1989gv}

Thus, the PT one-loop gluon self energy is to be identified with the term $\widehat{\Pi}^{(1)}_{\alpha\beta}$  {\it alone}; and one has in addition that  the modified gluon and ghost PT self-energy contributions are individually transverse\footnote{An explicit two-loop verification of the fat that PT gluon and ghost loops are individually transverse and that loops of different order do not mix is provided in~\cite{Binosi:2009qm}, see in particular Figs. 20 and 77 and Eqs.~(3.40) and (3.41). The all-order proof has been presented in~\cite{Aguilar:2006gr}.}: 
\begin{equation}
	q^\beta\widehat{\Pi}^{\mathrm{gl},(1)}_{\alpha\beta}(q)=0;\qquad  q^\beta\widehat{\Pi}^{\mathrm{gh},(1)}_{\alpha\beta}(q)=0.
	\label{newtrans}
\end{equation}
The one-loop PT self-energy $\widehat{\Pi}^{(1)}_{\alpha\beta}$ may be further evaluated, using the results\footnote{I use the momenta routing defined in~\1eq{1loopglseconv}.}
\begin{align}
	\widetilde{\Gamma}^{(0)}_{\alpha \mu\nu}\widetilde{\Gamma}^{(0)\mu\nu}_{\beta}= d (2k+q)_{\alpha} (2k+q)_{\beta} + 8 q^2 P_{\alpha\beta}(q),
\end{align}
and
\begin{align}
	\int_k\!\frac{\widetilde{\Gamma}^{(0)}_{\alpha}\widetilde{\Gamma}^{(0)}_{\beta}}{k^2 (k+q)^2}=\int_k \frac{(2k+q)_{\alpha} (2k+q)_{\beta}}{k^2 (k+q)^2} = - \left(\frac{1}{d-1}\right) q^2 P_{\alpha \beta}(q)\int_k \frac{1}{k^2 (k+q)^2},
\end{align}
to finally cast it in the simple form
\begin{align}
\widehat{\Pi}^{(1)}_{\alpha \beta}(q) = \left(\frac{7d-6}{d-1}\right) g^{2}\frac{C_A}{2} 
q^2 P_{\alpha \beta}(q) \int_k \frac{1}{k^2 (k+q)^2}.
\label{Pid}
\end{align}
Writing finally $\widehat{\Pi}^{(1)}_{\alpha \beta}= \widehat\Pi^{(1)}P_{\alpha \beta}$ and following the standard integration rules for the Feynman integral, I obtain the unrenormalized $\widehat\Pi^{(1)}$:
\begin{align}
	\widehat\Pi^{(1)}(q^2) &= ib g^2 q^2 
\left[\frac{2}{\epsilon} + \ln 4\pi -\gamma_{ E}  - \ln \frac{q^2}{\mu^2} + \frac{67}{33}\right];& b&= \frac{11C_{A}}{48\pi^2},
\label{1l_PT_prop_res}
\end{align}
where $b$ is the one-loop coefficient of QCD's $\beta$ function ($\beta = -bg^3$) in the absence of quark loops, and $\gamma_{ E}$ is the Euler-Mascheroni constant ($\gamma_{E}\approx0.57721$).

The appearance of $b$ in front of the logarithm is not accidental, and is exactly what happens with the vacuum polarization of QED. In the latter case the corresponding coefficient is negative, the difference in the sign being related to the fact that QCD is asymptotically free while QED is not. The fact that the PT gluon propagator captures the leading Renormalization Group (RG) logarithms is a direct consequence of the fact that, exactly as in QED again, the one-loop charge renormalization constant, $Z^{(1)}_g$, and the wave-function renormalization of the PT gluon self-energy, $Z^{(1)}_B$, are related by $Z^{(1)}_{g}=1/\sqrt{Z_B^{(1)}}$, so that the combination 
\begin{align}
	\widehat{d}^{(1)}(q^2)&=\alpha_s(\mu^2)\widehat{\Delta}^{(1)}(q^2;\mu^2);& \alpha_s(\mu^2)&=\frac{g^2(\mu^2)}{4\pi},
	\label{effchrg}
\end{align} 
is a Renormalization Group (RG) invariant combination. We will soon see that this one-loop result is in fact an all-orders one.

\paragraph{Background Field Method.}

\begin{figure}[!t]
	\centering
	\includegraphics[scale=0.356]{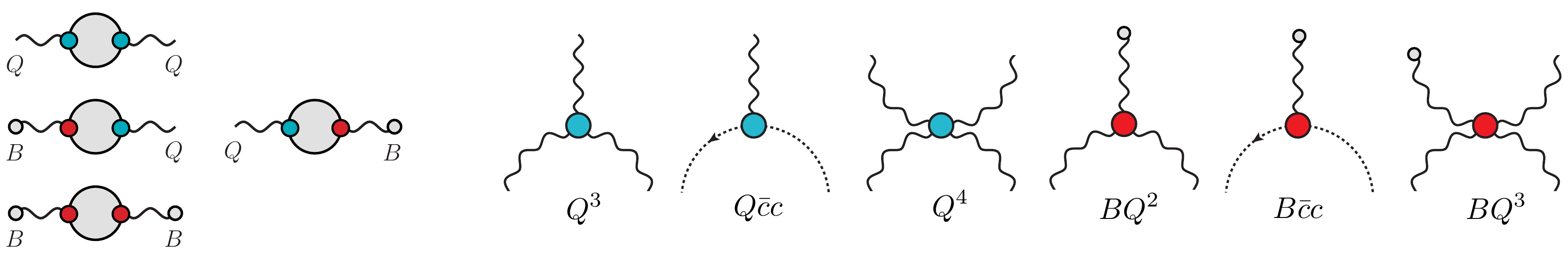}
	\caption{\label{BFM}Proliferation of gluon 2-, 3- and 4-point Green functions due to the splitting of the gluon field ($A$) into a background ($B$) and a quantum fluctuating part ($Q$).}
\end{figure}

Using the same type of construction outlined above the two-loop PT gluon self-energy was constructed in~\cite{Papavassiliou:1999az}. However, it is clear that going past the two-loop level requires transcending the PT diagrammatic origin and developing it as a fully-fledged algebraic technique. A step towards this goal is provided by the Background Field Method (BFM) gauge~\cite{Abbott:1980hw,Abbott:1981ke}, which provides the necessary Feynman rules to construct the PT gluon self-energy, as well as all other PT $n$-point Green functions. Indeed, as was shown in~\cite{Binosi:2009qm,Binosi:2002ft,Binosi:2003rr}, one has the important all-order Green functions equality
\begin{align}
	\Gamma^n_{\rm{PT}}\equiv\Gamma^n_{\rm{BFM}}.
\end{align}

The BFM $R_\xi$ gauge can be obtained by splitting the gauge field into a background ($B$) and a (quantum) fluctuating part ($Q$) according to
\begin{align}
	A^a_\mu=B^a_\mu+Q^a_\mu,
	\label{split}
\end{align} 
and demanding that the gauge fixed Lagrangian has a residual gauge invariance with respect to the $B$ field. This can be achieved by choosing a gauge-fixing function transforming in the adjoint representation of SU($N$), that is through the following replacements in~\1eq{lincov}
\begin{align}
	\partial_\mu\delta^{ab}\to\widetilde{\cal D}^{ab}_\mu\equiv\partial_\mu\delta^{ab}+f^{acb} B^c_\mu;\qquad
A^a_\mu\to Q^a_\mu.
\end{align}
Once implemented, I arrive at the BFM $R_\xi$ gauges fixing function
\begin{align}
	\widetilde{\cal F}^a=\widetilde{\cal D}^{ab}_\mu Q^\mu_b.
	\label{BFM-gf}
\end{align}
Inserting~\1eq{BFM-gf} into~\1eq{GF+FPG} one obtains the Feynman rules characteristic of the BFM, namely a symmetric $B\bar cc$ trilinear vertex and the four particle vertex $BQ\bar cc$. Inserting finally~\1eq{split} back into the original invariant Lagrangian, one gets the conventional Feynman rules together with those involving the background fields $B$; however, to lowest order, only vertices containing exactly two quantum fields $Q$ will differ from the conventional ones, see~\fig{BFM}. 

As a result of the residual gauge invariance, the contraction of Green functions with the momentum corresponding to a background gluon $B$ leads to Abelian-like Slavnov-Taylor identities (STIs), that is, linear identities that preserve to all orders their tree-level form\footnote{The divergence of a quantum field $Q$ gives rise instead to non-Abelian STIs, akin to those obtained in the conventional $R_\xi$ gauges.}. For example\footnote{I will indicate with a `tilde' (respectively, a `hat') quantities involving a single (respectively, two or more) background field(s). The notation used in Sect.~\ref{1lPT} should now be clear(er).}, the $BQ^2$ vertex $\widetilde{\Gamma}_{\mu\alpha\beta}$ and the $B\bar cc$ vertex $\widetilde{\Gamma}_{\mu}$, whose tree-level version has already appeared during our explicit construction of the one-loop PT self-energy in~\2eqs{GF}{Bcbc} respectively, satisfy the linear STIs (omitting color and assuming all momenta entering)
\begin{subequations}
\begin{align}
	q^\mu \widetilde{\Gamma}_{\mu\alpha\beta}(q,r,p) &= i\Delta_{\alpha\beta}^{-1}(r) - i\Delta_{\alpha\beta}^{-1}(p),\label{AbWI3gl}\\
	q^\mu \widetilde{\Gamma}_\mu(q,r,p) &= iD^{-1}(r^2) - iD^{-1}(p^2),\label{AbWI3gh}
\end{align}
\label{AbWI3}
\end{subequations}
where for a general $\xi$ I define\footnote{I omit the explicit $\xi$ dependence from the arguments of $\Delta$, $D$ and $F$.}
\begin{subequations}
	\begin{align}
	i\Delta_{\alpha\beta}(q)&=-i\left[P_{\alpha\beta}(q)\Delta(q^2)+\xi \frac{q_\alpha q_\beta}{q^4}\right];& \Delta(q^2)&=\frac1{q^2+i\Pi(q^2)},&\label{gl-prop}\\
	\Delta_{\alpha\beta}^{-1}(q)&=i\left[P_{\alpha\beta}(q)\Delta^{-1}(q^2)+\frac{i}{\xi}q_\alpha q_\beta\right];&	
	D^{-1}(q^2)&=-iq^2F^{-1}(q^2),
\end{align}
\label{invglpropdef}
\end{subequations}
\hspace{-0.18cm} with $F$ the so-called ghost dressing function. Additionally, the $BQ^3$ vertex satisfies
\begin{align}
    q^\mu \widetilde{\Gamma}^{mnrs}_{\mu\alpha\beta\gamma}(q,r,p,t) &= f^{mse}f^{ern} \Gamma_{\alpha\beta\gamma}(r,p,q+t) + f^{mne}f^{esr}\Gamma_{\beta\gamma\alpha}(p,t,q+r) \nonumber \\
	&\mathrel{\phantom{=}}+ f^{mre}f^{ens} \Gamma_{\gamma\alpha\beta}(t,r,q+p),
	\label{AbWI4gl}
\end{align}
with $\Gamma_{\alpha\mu\nu}$ the $Q^3$ three gluon vertex (see \1eq{tgv_ch2} for its tree-level definition).
These identities ensure that the two-point mixed background/quantum as well as the background/background gluon functions are such that the one- and two-loop dressed gluon and ghost diagrams subsets contributing to them are individually transverse (when contracted with the momentum of a background leg). \1eq{newtrans} represents an explicit one-loop proof of this fact. The transversality property of subsets of diagrams is of particular interest to DSE practitioners, as it allows to develop truncation schemes that preserve the transversality of the answer even if entire classes of diagrams are left out~\cite{Binosi:2007pi,Binosi:2008qk}. In addition, the residual gauge invariance ensures  the all-order relation $Z_{g}=1/\sqrt{Z_B}$ where $Z_g$ is the charge renormalization constant and $\sqrt{Z_B}$ the one of the $B$ fields; thus, as anticipated, the combination 
\begin{align}
	\widehat{d}(q^2)&=\alpha_s(\mu^2)\widehat{\Delta}(q^2;\mu^2),	
	\label{ao-effchrg}
\end{align} 
is RG invariant to all orders in perturbation theory.

\begin{figure}[!t]
	\centering
	\includegraphics[scale=0.75]{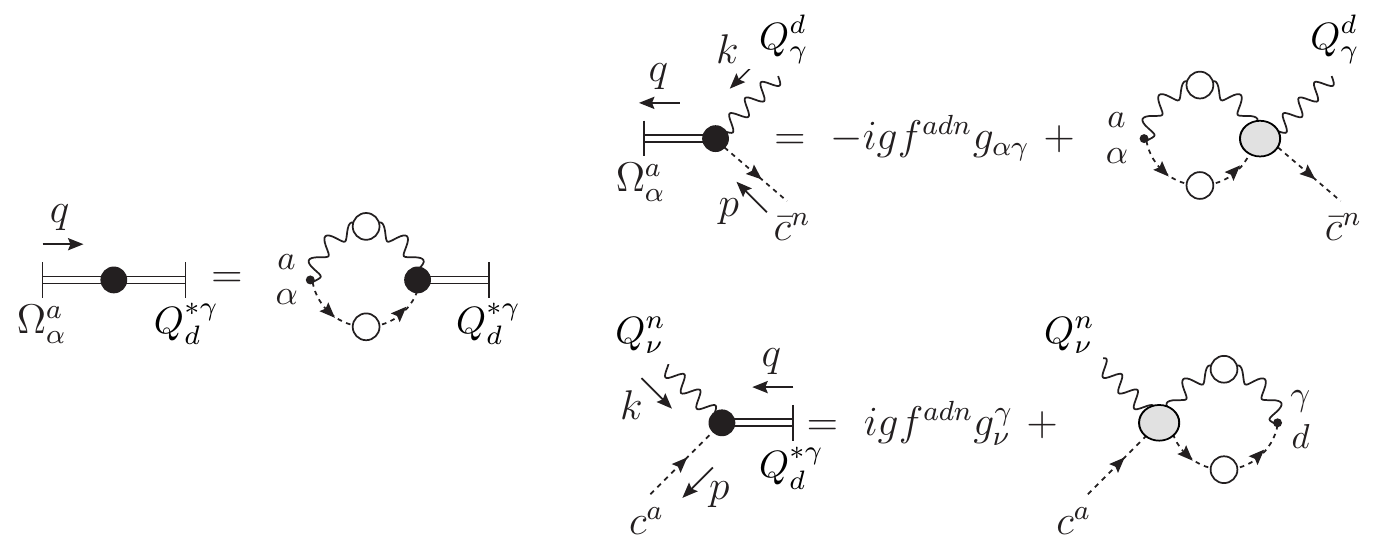}
	\caption{\label{fig:G-fig}Diagrammatic expansion of the auxiliary function $\Gamma_{\Omega_\alpha^a Q^{*\gamma}_d}$.}	
\end{figure}

\paragraph{Background Quantum Identities.} It turns out that the conventional and BFM $R_\xi$ gauges are related by symmetry transformations. In fact, as shown in~\cite{Binosi:2013cea}, Yang-Mills theories quantized in the BFM emerge in a natural way from Yang-Mills theories quantized in the $R_\xi$ gauges if one renders the latter invariant also under the anti-BRST symmetry. This is a crucial construction, because it clarifies the origin  of a plethora of identities, among which there are the so-called Background-Quantum identities (BQIs)~\cite{Binosi:2002ez,Grassi:1999tp}, relating (to all orders) Green functions evaluated in the conventional $R_\xi$ gauges to the same functions evaluated in the BFM $R_\xi$ gauges. The simplest of these identities, namely the one connecting the corresponding gluon propagators\footnote{A one-loop version of this identity was explicitly constructed in~\1eq{propexp}.}, turns out to be of paramount importance for the ensuing analysis. They read
\begin{subequations}
\begin{align}
i\Gamma_{B_\alpha^a    Q_\beta^b}(q)&=\left[ig_\alpha^\gamma
\delta^{ad}+ \Gamma_{\Omega_\alpha^a
Q^{*\gamma}_d}(q)\right]\Gamma_{Q^d_\gamma Q^b_\beta}(q),
\label{twoBQI1}\\
i\Gamma_{B_\alpha^aB_\beta^b}(q)&=\left[i g_\alpha^\gamma
\delta^{ad}+ 
\Gamma_{\Omega_\alpha^a Q^{*\gamma}_d}(q)\right]\Gamma_{Q^d_\gamma
B^b_\beta}(q),
\label{twoBQI2}
\end{align}
\label{twoBQI}
\end{subequations}
\hspace{-0.18cm} where
\begin{subequations}
\begin{align}
	\Gamma_{Q^a_\alpha Q^b_\beta}(q)&= i\delta^{ab} P_{\alpha\beta}(q) \Delta^{-1}(q^2),\label{gainvprop} \\
	\Gamma_{B^a_\alpha Q^b_\beta}(q)&= \Gamma_{Q^a_\alpha B^b_\beta}(q)=i\delta^{ab} P_{\alpha\beta}(q) \widetilde{\Delta}^{-1}(q^2), \\
	\Gamma_{B^a_\alpha B^b_\beta}(q)&= i\delta^{ab} P_{\alpha\beta}(q) \widehat\Delta^{-1}(q^2),
\end{align} 
\end{subequations} 
and $\Gamma_{\Omega_\mu^m Q^{*n}_\nu}$ is a special Green function capturing the gluon ghost dynamics with\footnote{In this auxiliary Green function $Q^*$ represents the gauge boson anti-field characteristic of the Batalin-Vilkovisky formulation of gauge theories~\cite{Batalin:1977pb,Batalin:1983ggl}; $\Omega$ is instead a classical source which forms a BRST doublet with the background field $B$ and is used to implement the equation of motions of the background fields~\cite{Grassi:1999tp,Grassi:2001zz}.}
\begin{align}
	\Gamma_{\Omega_\alpha^a Q^{*\gamma}_d}(q)&=\delta^{ad}G(q^2)g_\alpha^\gamma+\delta^{ad}L
	(q^2)\frac{q_\alpha q^\gamma}{q^2},
\end{align}
and the diagrammatic expansion given in Fig.~\ref{fig:G-fig}. The form factors $G$ and $L$ are related, due to antiBRST symmetry, to the ghost (inverse) dressing function through\footnote{This relation is true in the Landau gauge; for a general gauge fixing parameter $\xi$ it acquires an additional term~\cite{Binosi:2013cea}.}
\begin{align}
	F^{-1}(q^2)=1+G(q^2)+L(q^2).
	\label{antighost-rel}
\end{align}
Since $L(0)=0$~\cite{Aguilar:2009nf} one has $F^{-1}(0)=1+G(0)$ which identifies~\cite{Aguilar:2009pp}  $G$ as the Kugo-Ojima function~\cite{Kugo:1979gm}.

we can now combine Eqs.~(\ref{twoBQI}) such that
the two-point function mixing background and quantum fields drops out, to get the BQI
\begin{align}
i\Gamma_{B_\alpha^aB_\beta^b}(q)&=i\Gamma_{Q^a_\alpha Q^b_\beta}(q)+\Gamma_{\Omega_\alpha^a Q^{*\gamma}_d}(q)\Gamma_{Q^d_\gamma Q^b_\beta}(q)+\Gamma_{\Omega_\beta^b Q^{*\gamma}_d}(q)\Gamma_{Q^a_\alpha Q^d_\gamma}(q)\nonumber \\
&\mathrel{\phantom{=}}-i\Gamma_{\Omega_\alpha^a A^{*\gamma}_d}(q)\Gamma_{Q^d_\gamma Q^{e}_{\epsilon}}(q)\Gamma_{\Omega_\beta^b Q^{*\epsilon}_{e}}(q)\nonumber \\
&=i\Gamma_{Q^a_\alpha Q^b_\beta}(q)+2\Gamma_{\Omega_\alpha^a Q^{*\gamma}_d}(q)\Gamma_{Q^d_\gamma Q^b_\beta}(q)\nonumber \\
&\mathrel{\phantom{=}}-i\Gamma_{\Omega_\alpha^a Q^{*\gamma}_d}(q)\Gamma_{Q^d_\gamma Q^{e}_{\epsilon}}(q)\Gamma_{\Omega_\beta^b Q^{*\epsilon}_{e}}(q),
\label{BQI:gg}
\end{align}
where the last identity is due to the transversality of the gluon two-point function.
At one loop
\begin{align}
	\Gamma^{(1)}_{B_\alpha^aB_\beta^b}(q)&=\Gamma^{(1)}_{Q^a_\alpha Q^b_\beta}(q)-2i\Gamma^{(1)}_{\Omega_\alpha^a Q^{*\gamma}_d}(q)\Gamma^{(0)}_{Q^d_\gamma Q^b_\beta}(q),
\end{align}
and using the relations~\noeq{gainvprop}, as well as the expression
\begin{subequations}
\begin{align}
	\Gamma^{(1)}_{\Omega_\alpha^a Q^{*\gamma}_d}(q)&=-g^2C_{A}\delta^{ad}g^\gamma_\alpha\int_k\!\frac{1}{k^2 (k+q)^2},
\end{align}	
\end{subequations}
one obtains
\begin{align}
	\widehat{\Pi}^{(1)}_{\alpha\beta}(q)=\Pi^{(1)}_{\alpha\beta}(q)+2\Pi^{P,(1)}_{\alpha\beta}(q),
\end{align}
{\it i.e.}, the same relation obtained in~\noeq{propexp} as anticipated. Thus the PT can be viewed as a diagrammatic way to expose BQIs; and, as a result,  we can work directly with these identities. This framework is now known as PT-BFM.    

To conclude, observe that, using the transversality of the gluon 2-point functions, the BQIs of Eqs.~(\ref{twoBQI}) can be cast in the more compact form
\begin{align}
	\Delta(q^2)&=[1+G(q^2)]\widetilde{\Delta}(q^2);&
	\widetilde{\Delta}(q^2)&=[1+G(q^2)]\widehat\Delta(q^2),
	\label{BQIf}
\end{align}
which will finally allow to connect the RG invariant combination $\widehat{d}$ of~\1eq{ao-effchrg} with the conventional propagator $\Delta$:
\begin{align}
	\widehat{d}(q^2)=\frac{\Delta(q^2;\mu^2)}{\left[1+G(q^2;\mu^2)\right]^2}.
	\label{ec-start}
\end{align}
 
\paragraph{Why the gluon should be massless...}

\begin{figure}[!t]
    \centering
	\includegraphics[scale=.4]{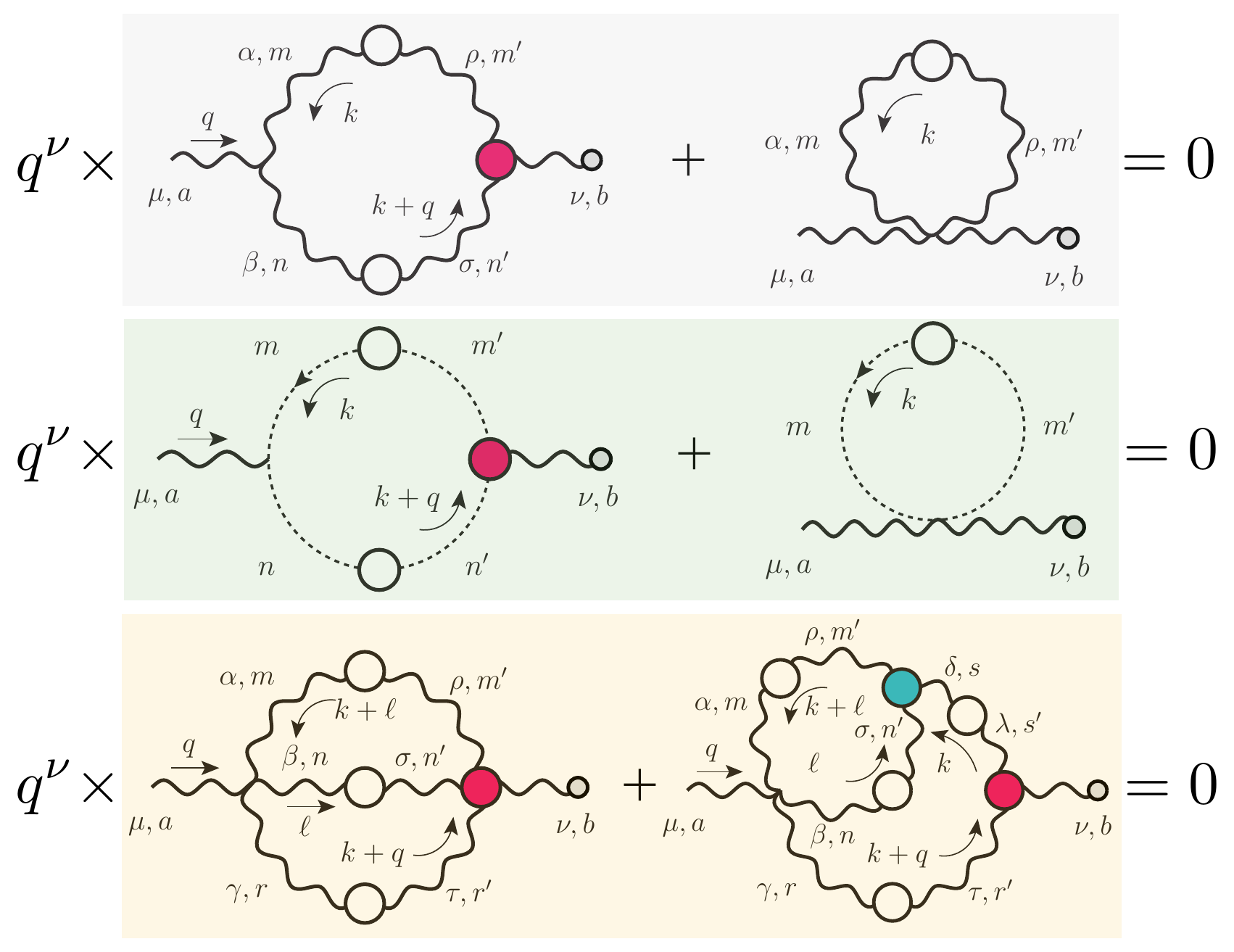}	
	\caption{\label{fig:QB-DSE-and-STI}From top to bottom: one-loop dressed gluon and ghost, and two-loop dressed gluon contributions to the mixed $QB$ gluon DSE. Also highlighted is the fact that the divergence of each individual subset vanishes as required by background gauge invariance.}
\end{figure}

Before we can understand how the gluon can acquire a dynamically generated mass, we first need to get acquainted with the mechanism that forbids it to acquire one. Of course, gauge invariance does not allow to add an {\it explicit} gluon mass term in the QCD Lagrangian~\noeq{lagden}; but it does not constrain what dynamics can (gauge-invariantly) do. The reason why even dynamically $m^2_{\mathrm{gl}}=0$, is easier to grasp in the PT-BFM framework, where one can consider the (Landau gauge) mixed quantum/background ($QB$) self-energy $\widetilde\Pi_{\alpha\beta}$, with its DSE given in Fig.~\ref{fig:QB-DSE-and-STI}. Since the latter is related to the conventional $QQ$ propagator through the BQI~\noeq{BQIf}, one has
\begin{align}
	\Delta^{-1}(q^2)P_{\alpha\beta}(q)=\frac{q^2P_{\alpha\beta}(q)+\widetilde\Pi_{\alpha\beta}(q)}{1+G(q^2)},
	\label{QQtoBQ}
\end{align}  
and therefore, any conclusion reached in for the $QB$ case directly translates to the conventional propagator (and the background $BB$ as well). One has
\begin{align}
	\widetilde\Pi_{\alpha\beta}(q)&=\widetilde\Pi^{\rm{gl}}_{\alpha\beta}(q)+\widetilde\Pi^{\rm{gh}}_{\alpha\beta}(q)+\widetilde\Pi^{\rm{gl;2l}}_{\alpha\beta}(q),
\end{align}
with
\begin{subequations}
	\begin{align}
		\widetilde\Pi^{\rm{gl}}_{\alpha\beta}(q)&=-\frac{g^2}2C_A\hspace{-0.11cm}\int_k\!\Gamma^{(0)}_{\mu\alpha\beta}(q,k,-k-q)\Delta^{\alpha\rho}(k)\Delta^{\beta\sigma}(k+q)\widetilde{\Gamma}_{\nu\sigma\rho}(-q,k+q,-k)\nonumber \\
		&\mathrel{\phantom{=}}+g^2C_A\hspace{-0.11cm}\int_k\!\left[g_{\mu\nu}\Delta^\alpha_\alpha(k)-\Delta_{\mu\nu}(k)\right],\label{1lgl}\\
		\widetilde\Pi^{\rm{gh}}_{\alpha\beta}(q)&=-g^2C_A\hspace{-0.11cm}\int_k\! k_\mu D(k)D(k-q)\widetilde\Gamma_\nu(-q,k,-k+q)\nonumber \\
		&\mathrel{\phantom{=}}+g^2C_Ag_{\mu\nu}\hspace{-0.11cm}\int_k\! D(k),\label{1lgh}\\
		\widetilde\Pi^{\rm{gl;2l}}_{\alpha\beta}(q)&=i\frac{g^4}6\Gamma^{(0)amnr}_{\mu\alpha\beta\gamma}\hspace{-0.15cm}\int_{k,\ell}\hspace{-0.2cm}\Delta^{\alpha\rho}(k+\ell)\Delta^{\beta\sigma}(\ell)\Delta^{\gamma\tau}(k+q)\nonumber \\
		&\mathrel{\phantom{=}}\times\widetilde{\Gamma}_{\nu\tau\sigma\rho}^{brnm}(-q,k+q,\ell,-\ell-k)\nonumber \\
		&\mathrel{\phantom{=}}+i\frac{g^4}2f^{bre}f^{emn}\Gamma^{(0)amnr}_{\mu\alpha\beta\gamma}\hspace{-0.11cm}\int_k\!Y^{\alpha\beta}_{\delta}(k)\Delta^{\delta\lambda}(k)\Delta^{\gamma\tau}(k+q)\nonumber \\
		&\mathrel{\phantom{=}}\times \widetilde{\Gamma}_{\nu\tau\lambda}(-q,k+q,-k),\label{2lgl}
	\end{align}
	\label{glDSEterms}
\end{subequations}
\hspace{-0.18cm} and
\begin{align}
	Y^{\alpha\beta}_{\delta}(k)&=\hspace{-0.11cm}\int_\ell\!\Delta^{\alpha\rho}(k+\ell)\Delta^{\beta\sigma}(\ell)\Gamma_{\sigma\rho\delta}(\ell,-k-\ell,k)\nonumber \\
	&=(g^\beta_\delta k^\alpha-g^\alpha_\delta k^\beta)Y(k^2).
	\label{defY}
\end{align}

Let me then {\it assume} that vertices carrying the $B$ leg do not contain massless poles of the type $1/q^2$, so that one can Taylor expand  both sides of~\1eq{AbWI3} around $q^2=0$ to get
\begin{subequations}
\begin{align}
    &\widetilde{\Gamma}_{\mu\alpha\beta}(0,r,-r) = -i \frac{\partial }{\partial r^\mu}\Delta^{-1}_{\alpha\beta}(r),\label{WTI3gl0}\\
    &\widetilde{\Gamma}_{\mu}(0,r,-r) = -i \frac{\partial }{\partial r^\mu}D^{-1}(r^2),
    \label{WTI3gh0}	
\end{align}
\label{WTI3}	
\end{subequations}
\hspace{-0.18cm} and, using~\1eq{AbWI4gl},
\begin{align}
    \widetilde{\Gamma}^{mnrs}_{\mu\alpha\beta\gamma}(0,-r,-p,r+p) &= -\left(f^{mne}f^{esr}\frac{\partial}{\partial r^\mu} + f^{mre}f^{ens}\frac{\partial}{\partial p^\mu}\right)\times\nonumber \\
    &\mathrel{\phantom{=}}\times\Gamma_{\alpha\beta\gamma}(-r,-p,r+p).
    \label{WTI4gl0}
\end{align}
Inserting these expressions in the gluon DSE terms~\noeq{glDSEterms}, yields after some algebra\footnote{At $q^2=0$ one has $\widetilde{\Pi}_{\alpha\beta}(0)\propto g_{\alpha\beta}$, and therefore it is sufficient to consider the self-energy trace only: $d\widetilde{\Pi}(0)=\widetilde{\Pi}^{\alpha}_{\alpha}(0)$.}
\begin{subequations}
	\begin{align}
	d\widetilde\Pi^{\rm{gl}}(0)&=\frac12g^2C_A\int_k\frac{\partial}{\partial k_\mu}[{\Gamma}^{(0)}_{\mu\alpha\beta}(0,k,-k)\Delta^{\alpha\beta}(k)],\label{1lgl0}\\
	d\widetilde\Pi^{\rm{gh}}(0)&=g^2C_A\int_k\frac{\partial}{\partial k_\mu}[{\Gamma}^{(0)}_{\mu}(0,k,-k)D(k)],\label{1lgh0}\\
	d\widetilde\Pi^{\rm{gl;2l}}(0)&=-i(d-1)g^4C^2_A\int_k\frac{\partial}{\partial k_\mu}[k_\mu\Delta(k^2)Y(k^2)],\label{2lgl0}
	\end{align}	
	\label{1lglall}
\end{subequations}
\hspace{-0.18cm} that is we have the final result
\begin{align}
    \widetilde{\Delta}^{-1}(0) & = \int_k\frac{\partial}{\partial k_\mu}{\cal F}_\mu(k);&
     {\cal F}_\mu(k) &=k_\mu\left\{ \Delta(k^2) \left[c_1 + c_2 Y(k^2)\right]+c_3D(k^2)\right\},
    \label{seag1}
\end{align}
with $c_1,c_2,c_3 \neq 0$. 

Now observe that since ${\cal F}_\mu$ is an odd function one has immediately
\begin{align}
	\int_k{\cal F}_\mu(k)=0.
	\label{intFvect}
\end{align}
Also, within dimensional regularization (or any other scheme that preserves translational invariance) one may shift the argument of the function ${\cal F}_\mu$ by an arbitrary momentum $q$ without compromising the result~\noeq{intFvect}. 
Then, carrying out a Taylor expansion around $q=0$, and using the result
\begin{align}
	{\cal F}_\mu (q+k) &= {\cal F}_\mu (k) + q^\nu\bigg\lbrace\frac{\partial }{\partial q^\nu}{\cal F}_\mu (q+k)\bigg\rbrace_{q=0} + {\cal O}(q^2)\nonumber \\
	&= {\cal F}_\mu (k) + q^\nu\frac{\partial {\cal F}_\mu (k)}{\partial k^\nu} + {\cal O}(q^2),
\label{TaylFvectq0}
\end{align} 
I obtain
\begin{align}
	q^\nu\int_k\frac{\partial {\cal F}_\mu (k)}{\partial k^\nu} = 0,
	\label{projectq}
\end{align}
since, in agreement with \1eq{intFvect}, if we integrate both sides of the above Taylor expansion, the result must vanish order by order. Given that the integral above has two free Lorentz indices and no momentum scale, it can only be proportional to the metric tensor $g_{\mu\nu}$; in addition, since $q$ is arbitrary, one concludes that~\1eq{projectq} is realized through the so-called `seagull identity'~\cite{Aguilar:2009ke,Aguilar:2016vin}
\begin{align}
\int_k\! \frac{\partial}{\partial k^\mu}{\cal F}_\mu (k) = 0,
\label{seagull}
\end{align}
which finally implies that the gluon remains rigorously massless:
\begin{align}
    \widetilde{\Delta}^{-1}(0) & =0\quad\Rightarrow\quad
    \Delta^{-1}(0) =0.
    \label{masslessness}
\end{align}

\paragraph{...and how it got its mass.}

This result may be circumvented by relaxing the assumption made when deriving \2eqs{WTI3}{WTI4gl0}, {\it i.e.}, I will now allow the vertices to contain longitudinally coupled $1/q^2$ poles. This is because, according to a fundamental observation made by Schwinger in the 60s~\cite{Schwinger:1962tn,Schwinger:1962tp}, if the dimensionless vacuum polarization $\Pi(q^2)=q^2\mathbf{\Pi}(q^2)$ develops a pole at zero momentum transfer ($q^2 = 0$), then the vector meson (gluon) acquires a mass. Indeed, if $\mathbf{\Pi}(q^2)= m^2_\mathrm{gl}/q^2$, then (in Euclidean space\footnote{In Euclidean space $p{\cdot}q=\sum_ip_iq_i$, and one has: $q^2_{\s{E}}=-q^2$; $\int_k=i\int_{k_{\s{E}}}$; and $\Delta_{\s{E}}(q_{\s{E}}^2)=-\Delta(-q^2_{\s{E}})$. Additionally, for quarks: $\{\gamma_\mu,\gamma_\nu\}=2\delta_{\mu\nu}$; $\gamma_\mu^\dagger=\gamma_\mu$.}) \1eq{invglpropdef} implies that 
\begin{align}
	\Delta^{-1}(0)=m^2_\mathrm{gl}>0,
\end{align}
thus evading the masslessness condition~\noeq{masslessness} and enabling the dynamical generation of a gluon mass~\cite{Jackiw:1973tr,Eichten:1974et,Poggio:1974qs,Smit:1974je}. 

\begin{figure}[!t]
    \centering
	\includegraphics[scale=0.42]{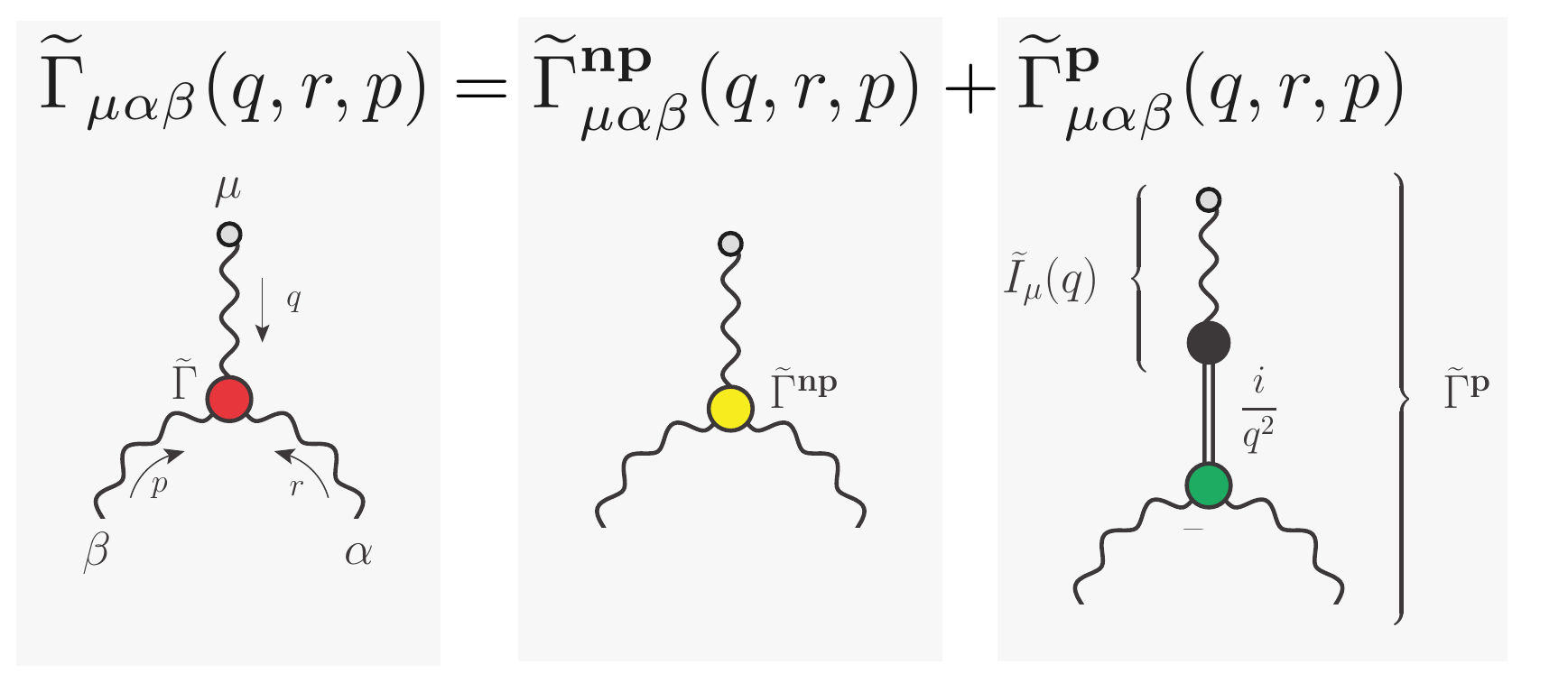}
	\caption{\label{fig:BQQ-np-p}The $BQ^2$ vertex decomposition into its regular (`{\bf np}') and pole (`{\bf p}') part. A similar relation holds for the ghost $B\bar c c$ vertex $\widetilde{\Gamma}_\mu$.}
\end{figure}

The dynamical realization of this concept at the level of a Yang-Mills theory requires the existence of a special class of nonperturbative vertices $\widetilde{C}$ (with appropriate color and Lorentz indices) that, when added to the conventional (fully dressed) vertices, have a triple effect: ({\it i}) they evade the seagull cancellation and cause the DSE of the gluon propagator to yield $\Delta^{-1}(0)>0$; ({\it ii}) they guarantee that the Abelian and non-Abelian STIs of the theory remain intact, {\it i.e.}, that they maintain exactly the same form before and after the mass generation; and ({\it iii}) they decouple from on-shell amplitudes. These crucial properties are possible because these special vertices: ({\it a}) contain massless poles and ({\it b}) are completely longitudinally coupled, {\it i.e.}, they satisfy conditions such as (for a three-gluon vertex)
\begin{align}
	P^{\alpha\alpha'}(q)P^{\mu\mu'}(r)P^{\nu\nu'}(p)\widetilde{C}_{\alpha'\mu'\nu'}(q,r,p)=0.
\end{align}
The origin of these  massless poles is due to purely non-perturbative dynamics: for sufficiently strong binding, the masses of certain (colored) bound states may be reduced to zero. Neglecting effects stemming from poles associated with the four-gluon vertex (which within the PT-BFM framework can be done in a gauge-invariant way), the $BQ^2$ and $B\bar cc$ vertices will then take the form (see~\fig{fig:BQQ-np-p})
\begin{subequations}
\begin{align}
    \widetilde{\Gamma}_{\mu\alpha\beta}(q,r,p) &= \g_{\mu\alpha\beta}(q,r,p) + i \frac{q_\mu}{q^2}\widetilde{C}_{\alpha\beta}(q,r,p),
    \label{GnpGp1}\\ 
    \widetilde{\Gamma}_{\mu}(q,r,p)&=\g_{\mu}(q,r,p)+ i \frac{q_\mu}{q^2}\Cgh(q,r,p), 
    \label{GnpGp2}
\end{align}
\end{subequations}
where the superscript ``np'' stands for ``no-pole'', whereas $\widetilde{C}_{\alpha\beta}$ and $\Cgh$ represents the bound-state gluon-gluon and gluon-ghost wave functions, respectively~\cite{Jackiw:1973tr,Eichten:1974et,Poggio:1974qs}.

Next, in order to preserve the BRST symmetry of the theory, 
I demand that all STIs maintain their exact form in the presence of these poles; therefore, Eqs.~(\ref{AbWI3}) will now read
\begin{subequations}
\begin{align}
	& q^\mu \g_{\mu\alpha\beta}(q,r,p) + \widetilde{C}_{\alpha\beta}(q,r,p) = i\Delta_{\alpha\beta}^{-1}(r) - i\Delta_{\alpha\beta}^{-1}(p),\label{glSTIwP}\\\
    & q^\mu\g_{\mu}(q,r,p) +\Cgh(q,r,p) = iD^{-1}(r^2) - iD^{-1}(p^2). 
    \label{ghSTIwP}
\end{align}
\label{STIwP}
\end{subequations}  
\hspace{-0.18cm} I now take the limit of~Eqs.~(\ref{STIwP}) as $q\to 0$ on both sides and match the different orders in $q$. The zeroth order yields the conditions
\begin{align}
    \widetilde{C}_{\alpha\beta}(0,r,-r)&=0;&
    \Cgh(0,r,-r)&=0,
    \label{zerothC}
\end{align}
whereas the first order furnishes a modified set of WTIs, namely
\begin{subequations}
\begin{align}
	&\g_{\mu\alpha\beta}(0,r,-r) = -i\frac{\partial}{\partial r^\mu}\Delta^{-1}_{\alpha\beta}(r) - \left\lbrace\frac{\partial}{\partial q^\mu}\widetilde{C}_{\alpha\beta}(q,r,-r-q)\right\rbrace_{q=0}, \label{WI2glwithpole}\\  
    &\g_\mu(0,r,-r)  = -i\frac{\partial}{\partial r^\mu} D^{-1}(r^2)- \left\lbrace\frac{\partial}{\partial q^\mu}\Cgh(q,r,-r-q)\right\rbrace_{q=0}.
    \label{WI2ghwithpole}
\end{align}
 \label{WI2new}	
\end{subequations}

The presence of the second term on the right-hand side of Eqs.~(\ref{WI2new}) has far-reaching consequences for the IR behavior of $\Delta(q^2)$. Specifically, a repetition of the steps leading to Eqs.~(\ref{1lglall}) and subsequently~\1eq{seag1}, reveals that, whereas the first terms on the right-hand side of these equations vanish again, the second terms survive, giving~\cite{Aguilar:2016vin} 
\begin{align}
    \Delta^{-1}(0)&=\frac32g^2C_AF(0)\bigg\{\int_k k^2 \Delta^2(k^2)\left[1-\frac32g^2C_AY(k^2)\right]\Cgl'(k^2)\nonumber \\
    &\mathrel{\phantom{=}}-\frac13\int_k k^2 D^2(k^2)\Cgh'(k^2)\bigg\},
    \label{DSEmass}
\end{align}
where $\Cgl$ is the form factor of $g_{\alpha\beta}$ in the tensorial decomposition of $\widetilde{C}_{\alpha\beta}$, and 
\begin{align}
    C_i^{\prime}(k^2)=\lim_{q\to0}\left\lbrace\frac{\partial \widetilde{C}_i(q,k,-k-q)}{\partial (k+q)^2}\right\rbrace,\quad i=\mathrm{gl},\ \mathrm{gh}.
\end{align}

As we see from \1eq{DSEmass}, a necessary condition for $\Delta^{-1}(0)$ to acquire a non-vanishing value is that at least one of the $\Cgl^{\prime}$ and $\Cgh^{\prime}$ does not vanish identically; in addition,  $\Cgl^{\prime}$ and $\Cgh^{\prime}$ must decrease sufficiently rapidly in the UV, in order for the integrals in \1eq{DSEmass} to give a (positive) finite value. In such a case, the non-vanishing of $\Cgl^{\prime}$ can be linked to the generation of a running gluon mass as it happens in the quark case (see Sect.~\ref{sect.quarks}). The IR saturation of the gluon propagator suggests in fact the physical parametrization  
\begin{align}
	\Delta^{-1}(q^2) =  q^2 J(q^2) + \hh (q^2),
\end{align}
where $\hh (0)> 0$, and for the `kinetic' part I have 
\begin{subequations}
	\begin{align}
		J(q^2;\mu^2)&\underset{q^2\ll\Lambda^2_{\s{\mathrm{QCD}}}}{\sim}\log q^2+{\cal O}(q^2),\\
		J(q^2;\mu^2)&\underset{q^2\gg\Lambda^2_{\s{\mathrm{QCD}}}}{\sim}[\log (q^2/\Lambda^2_{\s{\mathrm{QCD}}})/\log (\mu^2/\Lambda^2_{\s{\mathrm{QCD}}})]^\frac{\gamma_0}{\beta_0}\times[1+{\cal O}(q^2)],
	\end{align}
	\label{Jbehav}
\end{subequations}
\hspace{-0.18cm} with: $\Lambda_{\s{\mathrm{QCD}}}$ the mass-scale characterizing QCD perturbation theory; and $\gamma_0=13/2-2N_f/3$, $\beta_0=11-2N_f/3$ with $N_f$ the number of active quarks. Then the modified gluon STI~\noeq{glSTIwP} will make it natural to associate the $J$ terms with the $q^\mu\g_{\mu\alpha\beta}$ on the left-hand side, and, correspondingly, 
\begin{align}
	\widetilde{C}_{\alpha\beta}(q,r,p) &= \hh(p^2) P_{\alpha\beta}(p) - \hh(r^2) P_{\alpha\beta}(r).
    \label{thectilde}
\end{align}
Focusing on the $g_{\alpha\beta}$ components of \1eq{thectilde},
I obtain~\cite{Aguilar:2011xe} 
\begin{align}
\Cgl(q,r,p)=\hh(r^2)-\hh(p^2)\quad \underset{q\to0}{\Longrightarrow}\quad 
\Cgl'(r^2)=
\frac{\diff m^2(r^2)}{\diff r^2}.
\label{theCgl}
\end{align}    
Then, upon integration, I get
\begin{align}
    \hh(q^2)=\Delta^{-1}(0)+\int_0^{q^2}\!\!\diff y\,\Cgl^\prime(y),
    \label{Cglvsmass}
\end{align}
thus establishing the announced link between $\Cgl^\prime$ and a dynamically generated gluon mass~\cite{Binosi:2012sj}. However, in order for the quantity $\hh(q^2)$ to admit a running mass interpretation, it needs to: ({\it i}) be a monotonically decreasing function of $q^2$; ({\it ii}) vanish in the UV, {\it i.e.}, satisfy $\hh(\infty)=0$. In particular, the last condition implies
\begin{align}
	\Delta^{-1}(0)=-\int_0^{\infty}\!\!\diff y\,\Cgl^\prime(y).
	\label{glmass}
\end{align}

\paragraph{Massless poles BSE.}

\begin{figure}[!t]
    \centering
    \includegraphics[scale=0.425]{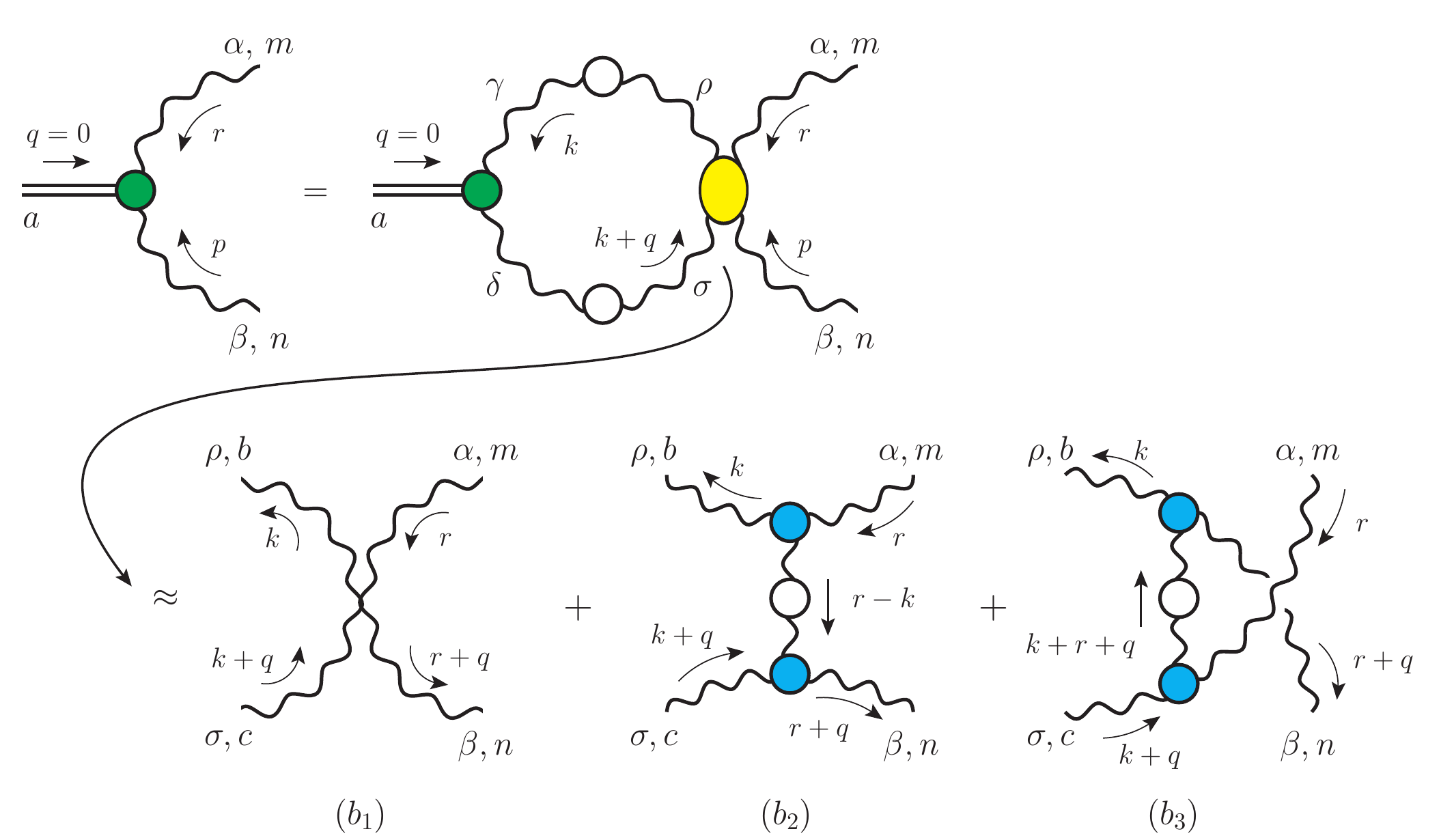}
    \caption{\label{fig:bse-4g}The BSE satisfied by the bound-state wave function $\widetilde{C}_{\alpha\beta}$ (upper line) and the simplified four gluon kernel used.}
\end{figure}

The actual dynamical realization of the scenario described in the previous subsection requires the study of the homogeneous Bethe-Salpeter equation (BSE) that controls the actual formation of the massless bound states to show that there are non-trivial solutions for $\Cgl'$ and $\Cgh'$. As it is sufficient to show that at least one of the BSE wave function is non-trivial, I will restrict my attention (again gauge invariantly) to the three gluon vertex case (see \fig{fig:bse-4g}), {\it i.e.}, show that $\Cgl'\neq0$.

The dynamical equation that governs $\Cgl'$ may be derived from the DSE satisfied by $\widetilde{\Gamma}_{\mu\alpha\beta}$ as $q \to 0$; in this limit, the derivative term becomes the leading contribution, given that as shown in~\1eq{zerothC} $\widetilde{C}_{\alpha\beta}(0,r,-r)=0$; and the resulting homogeneous equation assumes the form~\cite{Aguilar:2016ock}
\begin{align}
    f^{amn}\lim_{q\to 0} \widetilde{C}_{\alpha\beta}(q,r,p)
    &= f^{abc}\lim_{q\to 0} \bigg\{ \int_k \widetilde{C}_{\gamma\delta}(q,k,-k-q)\Delta^{\gamma\rho}(k)\nonumber \\
    &\mathrel{\phantom{=}}\times\Delta^{\delta\sigma}(k+q){\cal K}_{\rho\alpha\beta\sigma}^{bmnc}(-k,r,p,k+q)\bigg\}.
\label{BSEq}
\end{align}
Proceeding further requires to approximate in some way the four-gluon Bethe-Salpeter kernel ${\cal K}$; I will do that by considering the lowest-order set of diagrams appearing in its skeleton expansion, given by the diagrams $(b_1)$, $(b_2)$, and $(b_3)$, shown in the second line of \fig{fig:bse-4g}. It turns out that, if I use the tree-level four-gluon vertex in the evaluation of $(b_1)$, its contribution in the above kinematic limit vanishes. Diagrams $(b_2)$ and $(b_3)$, which carry a statistical factor of 1/2, are considered to contain fully dressed gluon propagators and $Q^3$ vertices $\Gamma$. As a consequence, the BSE~\noeq{BSEq} turns out to be renormalization group invariant~\cite{Binosi:2017rwj}.

The vertex $\Gamma$ contains 14 form factors~\cite{Ball:1980ax}, whose nonperturbative structure, albeit the subject of numerous studies~\cite{Alkofer:2008dt,Tissier:2011ey,Pelaez:2013cpa,Aguilar:2013vaa,Blum:2014gna,Eichmann:2014xya,Williams:2015cvx,Cyrol:2016tym}, is only partially known;  I shall then consider the simple Ansatz
\begin{align}
	\Gamma_{\mu\alpha\beta}(q,r,p)=
	\fQ (r)\Gamma^{(0)}_{\mu\alpha\beta}(q,r,p),
	\label{vertf}
\end{align} 
with $\fQ$ a suitable form factor that depends from a single kinematic variable. Then, substituting~\1eq{vertf} into \1eq{BSEq}, I arrive at the final equation~\cite{Binosi:2017rwj} 
\begin{align}
    \Cgl'(q^2)&=\frac{8\pi}3\alpha_s C_A\!\!\!\int_k\!\Cgl'(k^2)\frac{(q\!\cdot\!k)[q^2k^2-(q\!\cdot\! k)^2]}{q^4k^2(k+q)^2}\Delta^2(k)\Delta(k+q)\nonumber\\
    &\mathrel{\phantom{=}}\times \fQ^2(k+r)\left[8q^2k^2 + 6(q\!\cdot\!k)(q^2+k^2)+3(q^4+k^4)+(q\!\cdot\! k)^2\right].
    \label{masslessBSE}
\end{align}

\begin{figure}[!t]
	\centering
	\includegraphics[scale=0.55]{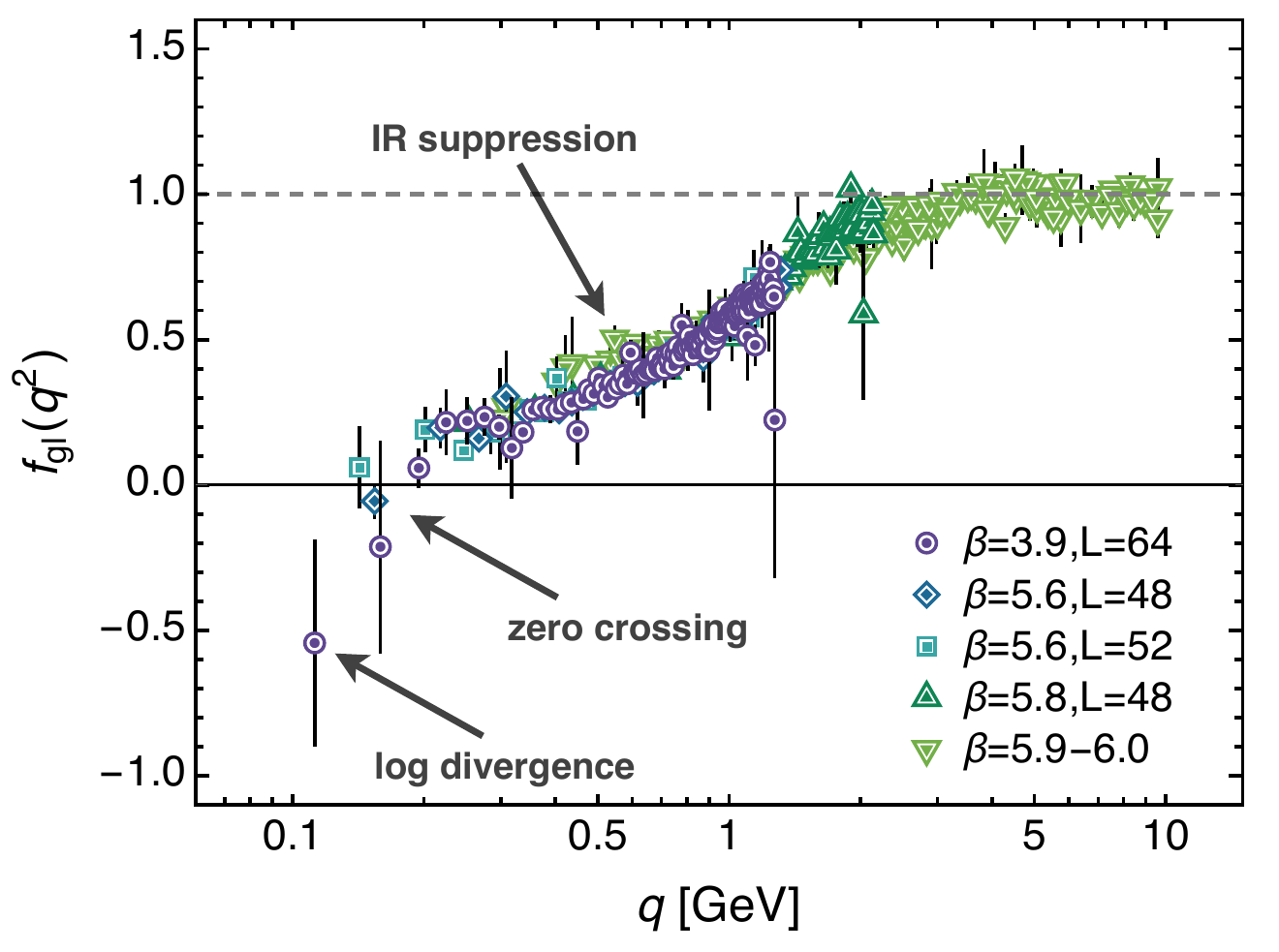}	
	\caption{\label{fig:fgl}Compilation of the latest lattice data for the SU(3) form factor $\fQ$ in the symmetric configuration. Notice: ({\it i}) the suppression at intermediate momenta with respect to the tree-level value $\fQ^{(0)}=1$; ({\it ii}) the zero-crossing; and ({\it iii}) the logarithmic divergence in the IR. All these features represents the signature of an IR massless ghost.}
\end{figure}

\begin{figure}[!t]
\centering
\includegraphics[scale=0.425]{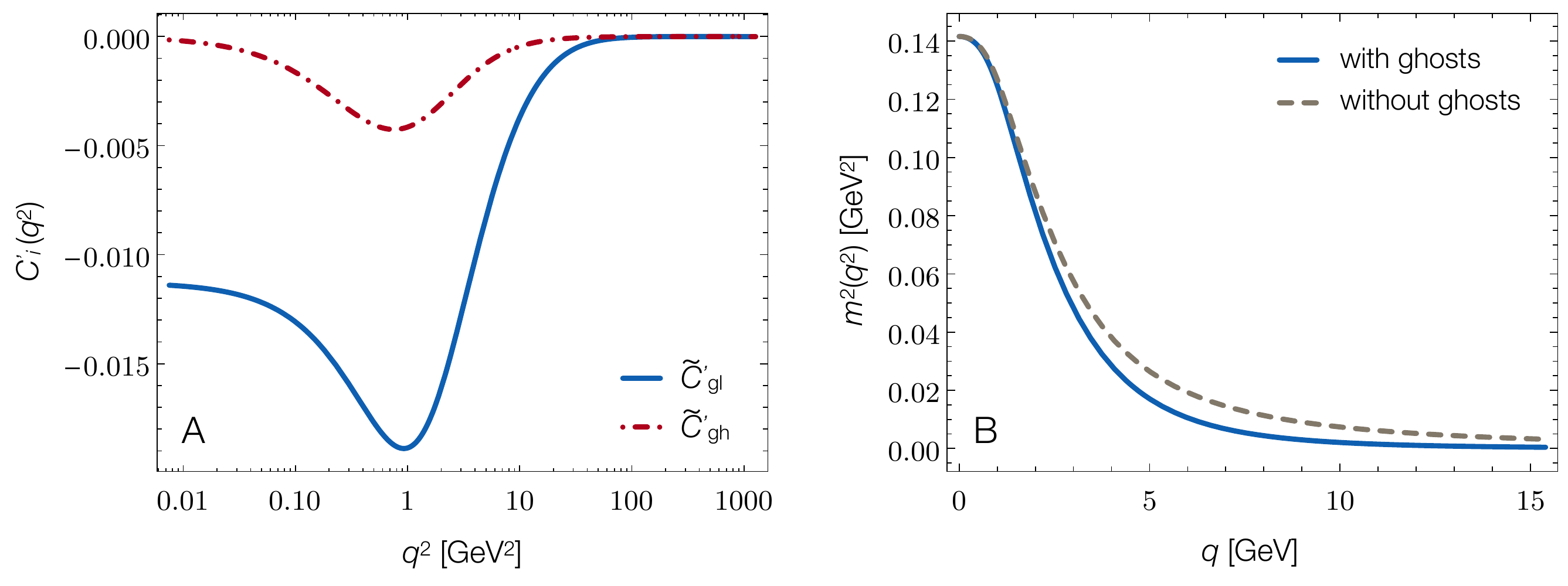}	
\caption{\label{fig:BSE-sols_and-m2} Normalized solution of the BSE~\noeq{masslessBSE} (panel A, blue continuous curve) and the corresponding running mass (panel B, dashed gray curve) obtained from the integration of~\1eq{Cglvsmass}. The red dot-dashed curve (respectively, blue continuous curve) in panel A (respectively, B)  show the ghost BSE amplitude (respectively, the running mass) when ghosts are added.}
\end{figure}

The functional form I will employ for $\fQ (r)$ is motivated by consistency requirements of the PT-BFM framework (see next section) and supported by a considerable number of studies in the continuum which have been confirmed on the lattice. In particular, for certain characteristic kinematic configurations (such as the symmetric and the soft gluon limits), the vertex is suppressed with respect to its tree-level value, reverses its sign for relatively small momenta (an effect known as ``zero crossing''), and finally diverges at the origin~\cite{Alkofer:2008dt,Tissier:2011ey,Pelaez:2013cpa,Aguilar:2013vaa,Blum:2014gna,Eichmann:2014xya,Williams:2015cvx,Cyrol:2016tym}. The reason for this particular behavior may be traced back to the delicate balance between contributions originating from gluon and ghost loops (see below). Early lattice verifications of the presence of a zero crossing in SU(2) Yang-Mills theories can be found in~\cite{Cucchieri:2006tf,Cucchieri:2008qm}, whereas the effect has recently been confirmed to be present also in the case of SU(3) theories~\cite{Athenodorou:2016oyh,Boucaud:2017obn,Duarte:2016ieu,Aguilar:2021lke}. The latest lattice data of~\cite{Aguilar:2021lke}, renormalized at $\mu=4.3$ GeV, are shown in~\fig{fig:fgl}. A typical (normalized) solution\footnote{Indeed~\1eq{glmass} shows that the normalization of the BSE solution must be negative.} of~\1eq{masslessBSE} is finally shown on panel A of~\fig{fig:BSE-sols_and-m2}, with the associated running gluon mass obtained from~\1eq{glmass} shown on panel B of the same figure. 

Let me now return to \1eq{DSEmass}; setting $\Cgh'=0$, I obtain a second order algebraic equation for $\alpha_s$, given by
\begin{align}
    A\alpha_s^2+B\alpha_s+C=0,
    \label{quadraticmass}
\end{align}
where, passing to Euclidean space, and using spherical coordinates,
\begin{subequations}
\begin{align}
    A&=\frac{3C^2_A}{32\pi^3}F(0)\hspace{-0.1cm}\int_0^\infty\hspace{-0.2cm}\diff y\, y^2{\Delta}^2(y) Y(y) \Cgl'(y), \\
    B&=-\frac{3C_A}{8\pi} F(0)\hspace{-0.1cm}\int_0^\infty\hspace{-0.2cm}\diff y\,  y^2{\Delta}^2(y) \Cgl'(y), \\
    C&=-\int_0^\infty\hspace{-0.2cm}\diff y\,\Cgl'(y).
	\label{ABC}
\end{align}
\end{subequations}
The unique positive solution of~\1eq{quadraticmass} is given by 
\begin{align}
    \alpha_s=\frac{-B+\sqrt{B^2-4AC}}{2A},
    \label{aDSE}
\end{align}	
which shows how the existence of a positive coupling relies on a delicate interplay between the strength of the one- and two-loop dressed contributions in the gluon DSE. To be sure, setting the three-gluon form factor to its tree-level value, {\it i.e.}, $\fQ=1$ both in \1eq{defY} and~\1eq{masslessBSE}, returns $\aDSE=0.42$ and $\aBSE=0.27$. It is only when the $\fQ$ matches the behavior shown in lattice simulations ({i.e.}, as $q^2$ decreases towards zero, $\fQ$ shows IR suppression, zero-crossing and log divergence), see~\fig{fig:fgl}, that one achieves the convergence of the couplings to the single value  $\aBSE=\aDSE=0.45$, which is reasonably closed, given the approximations employed, to the expected value of~$0.32$.

Finally, the impact of the ghost sector on the generation of the dynamical gluon mass was studied in~\cite{Aguilar:2017dco}. This implied assuming the presence of massless poles both in the $BQ^2$ three-gluon vertex as well as in the corresponding background gluon-ghost vertex~$B\bar c c$, and study the coupled system of the corresponding BSEs governing the dynamics of the amplitudes $\Cgl'$ and $\Cgh'$. The full analysis of the latter system reveals that (under suitable approximation of the gluon-ghost 4-particle kernel) the contribution of the poles associated with the ghost-gluon vertex are particularly suppressed (red dot-dashed curve in panel A of~\fig{fig:BSE-sols_and-m2}), their sole discernible effect being a slight modification in the running of the gluon mass, for momenta larger than a few GeV (blue continuous curve in  panel B of the same figure).

\subsection{Ghosts}

\begin{figure}[!t]
	\centering
	\includegraphics[scale=0.55]{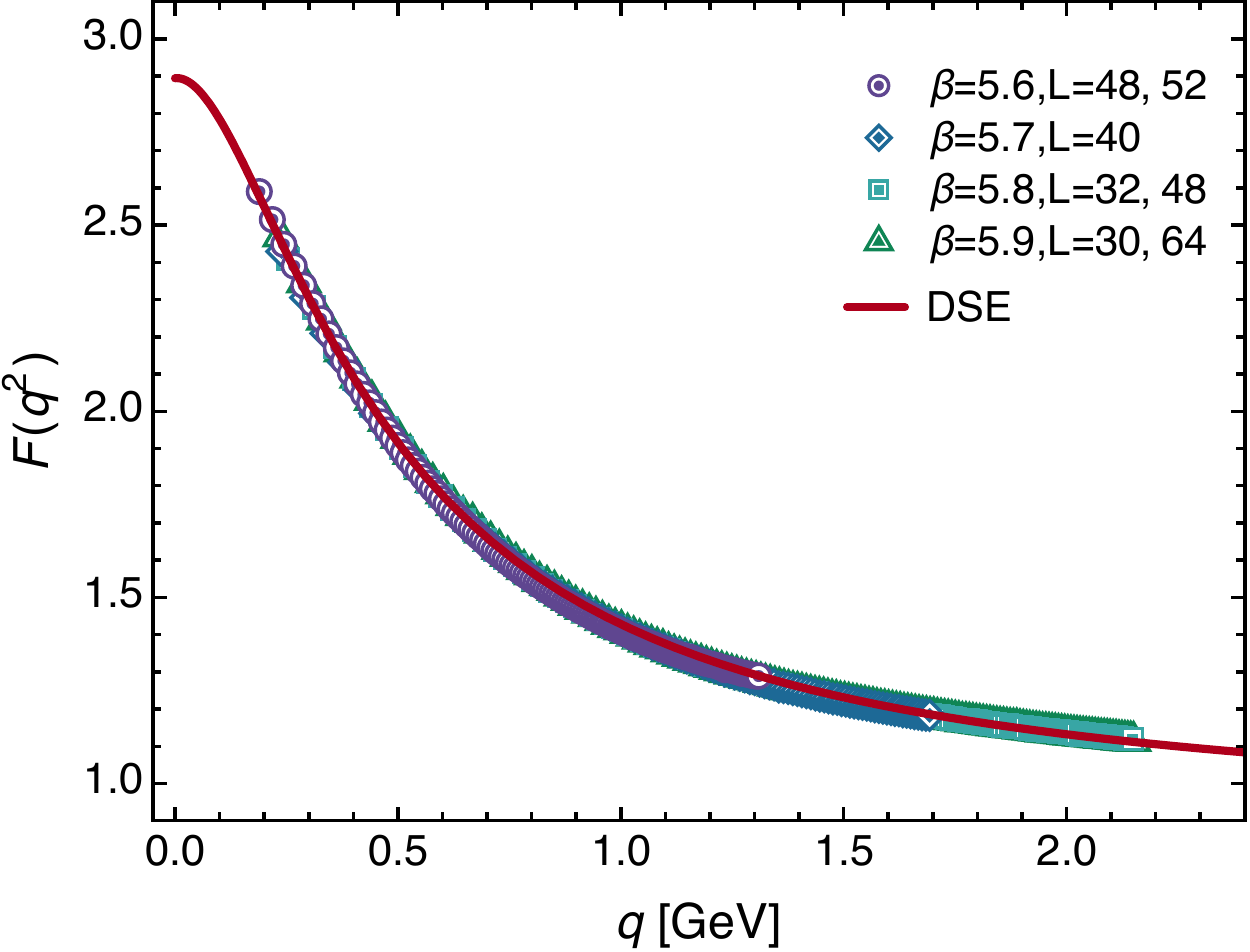}
	\caption{\label{fig:ghost-dress} Lattice data for the ghost dressing function $F$~\cite{Boucaud:2018xup} and the solution of the corresponding ghost DSE~\noeq{ghDSE} within the truncation scheme defined in~\cite{Aguilar:2021okw}.}
\end{figure}

Within the PT-BFM framework the dynamic generation of a gluon mass in the Landau gauge described in the previous section implies that also the ghost dressing function $F$, defined in the last of Eqs.~(\ref{invglpropdef}), saturates at a finite non-vanishing value; evidently \1eq{QQtoBQ} implies that\footnote{This result immediately invalidates the Kugo-Ojima  quartet mechanism for confinement, which would require \mbox{$G(0)=-1$} to work~\cite{Kugo:1979gm}. Notice that early lattice evidence that $G$ has never been close to reach -1 at $q^2=0$ can  already be found in~\cite{Sternbeck:2006rd}.} $G(0)\neq-1$, which inserted back to \1eq{antighost-rel} yields \mbox{$0<F^{-1}(0)<\infty$}. This prediction has been confirmed by lattice simulations that have shown that the ghost propagator behaves in the IR like $c/q^2$ so that $F(0)=c$ (see \fig{fig:ghost-dress}).

Therefore the ghost propagator, described by the DSE
\begin{align}
	D^{-1}(q^2)=q^2-g^2C_A\hspace{-0.11cm}\int_k\!(k+q)_\mu D(k+q)\Delta^{\mu\nu}(k)\Gamma_\nu(-k,k+q,-q),
	\label{ghDSE}
\end{align}
remains dynamically massless; and this dramatically affects many IR quantities. To begin with, the contribution to the kinetic part of the gluon propagator $J$ resulting from the ghost-loop diagram in the second line of \fig{fig:QB-DSE-and-STI} contains a pure  $\log q^2/\mu^2$ term without any mass in its argument possibly taming the associated IR divergence. This has to be contrasted  with the corresponding logs originating from gluon loops (first line of \fig{fig:QB-DSE-and-STI}), which read $\log[(q^2 + \hh)/\mu^2]$, and,  therefore are finite for arbitrary (Euclidean) momenta due to the explicit appearance of the dynamically generated gluon mass. Schematically, one has the relation~\cite{Aguilar:2013vaa}
\begin{align}
	J(q^2;\mu^2)\underset{q^2\to0}{\simeq}a+b\log\frac{q^2+\hh}{\mu^2}+c\log\frac
{q^2}{\mu^2},
\end{align}
and therefore, while the presence of the massless log does not interfere with the overall finiteness of $\Delta$ (simply because it is multiplied by $q^2$), its existence implies that: ({\it i}) the first derivative of the propagator diverges at the origin; and ({\it ii}) the gluon propagator is not a monotonic function of $q^2$, as it will display an IR maximum whose specific size and location is largely controlled by the relative weight of the massive and massless logs contributing to $J$. Both predictions have been confirmed by lattice simulations as can be seen in~\fig{fig:Delta-latt}; notice in particular that the position of the gluon propagator's maximum is in the deep IR, and its size is relatively small.

Now, in some special kinematic limits (which happens to be the ones typically studied on the lattice), the behavior of the three-gluon vertex can be predicted to be determined by the kinetic part of the gluon propagator $J$. This is the case, {\it e.g.}, when studying the ``symmetric'' configuration obtained setting in \1eq{tgv-glob} $k_1^2=k_2^2=q^2$ and $q{\cdot}k_1=q{\cdot}k_2=k_1{\cdot}k_2=-q^2/2$; then~\cite{Athenodorou:2016oyh}
\begin{align}
	\fQ(q^2;\mu^2)\simeq F(0;\mu^2)\frac\partial{\partial q^2}\left[\Delta^{-1}(q^2;\mu^2)\right],
\end{align} 
which shows that the dominant contribution as $q^2\to0$ is $\fQ\sim J\sim \log q^2$. Thus, for sufficiently small momenta $\fQ$ becomes negative, and then shows a log divergence at the origin\footnote{
In $d = 3$ the divergences induced due to the masslessness of the ghost are enhanced since in this case $J\sim1/q$. Correspondingly the maximum of the gluon propagator is clearly visible on the lattice~\cite{Cucchieri:2009xxr}, and so is the sudden negative divergence in $\fQ$~\cite{Cucchieri:2006tf}.}.
Notice finally that the sign of the divergence in $\fQ$ is a prediction of the PT-BFM framework too, being fixed by the sign of the (non-perturbative) log obtained from the massless ghost loop contribution to the gluon propagator in the IR. 

What has been described here is valid in the Landau gauge for both SU$(2)$ and SU$(3)$ gauge groups and quenched/unquenched configurations~\cite{Bogolubsky:2007ud,Bogolubsky:2009dc,Cucchieri:2010xr}. In the case of general $R_\xi$ gauges (with $\xi>0$) the situation is more complicated, as the tree-level term in the fully dressed gluon propagator proportional to the gauge-fixing parameter $\xi$, see~\1eq{gl-prop}, induces in the DSE~\noeq{ghDSE} a corresponding term that drives to zero the value of the dressing function at $q^2=0$~\cite{Huber:2015ria,Aguilar:2015nqa}. Additionally, whenever $\xi>0$ the form factor $\fQ$ ceases to be IR divergent and saturates at a negative $\xi$-dependent non-vanishing value, to ensure that the $R_\xi$ generalization of the massless poles BSE~\noeq{masslessBSE} still possesses acceptable solutions~\cite{Aguilar:2016ock}. It is currently a challenge for the lattice simulations to check these predictions at $\xi\neq0$. 

\subsection{Effective coupling\label{ssec:eff-coupl}}

We have now reached an excellent control over the gauge sector dynamics and the emergent phenomena that characterize the gluon and ghost propagator. To complete the study of this sector we need to construct a crucial quantity: the effective charge. The natural starting point in the PT-BFM framework is clearly provided by the RG invariant combination $\widehat{d}$ defined in~\1eq{ec-start}~\cite{Aguilar:2009nf,Binosi:2002vk}; then, recalling that $L(0)=0$, one immediately conclude that this quantity saturates at a finite non-vanishing values at $q^2=0$ owing to the dynamical generation of a gluon mass:
\begin{align}
	\widehat{d}(0)&=\frac{\alpha(\mu^2)}{\widehat{m}^2_{\mathrm{gl}}(\mu^2)}=\frac{\alpha_0}{m^2_0};& \widehat{m}^2_{\mathrm{gl}}(\mu^2)=\frac{m^2_{\mathrm{gl}}(0;\mu^2)}{F^2(0;\mu^2)}.
	\label{d0}
\end{align} 
As $\widehat{d}$ is a dimensionful quantity, to convert it into a dimensionless effective charge requires to factor out a (RG invariant) mass scale. To this end consider the RG invariant combination~\cite{Binosi:2016nme,Rodriguez-Quintero:2018wma,Cui:2019dwv}
\begin{align}
	{\mathrm D}(q^2)=\frac{1}{m_0^2} \Delta(q^2;\mu^2)\hh(0;\mu^2),
	\label{D}
\end{align}
and use~\1eq{Jbehav} to construct an interpolator ${\cal D}$ which accurately describes the available results for D on $q^2\lesssim\mu^2$ and is such that for $q^2\ll\Lambda^2_{\s{\mathrm{QCD}}}$ (respectively, $q^2\gg\Lambda^2_{\s{\mathrm{QCD}}}$) behaves like $m_0^2$ (respectively, $q^2$), and therefore represents in the far IR and UV the free propagator of a boson with mass $m_0$. Then one has the decomposition
\begin{align}
	\widehat{d}(q^2)=\widehat{\alpha}(q^2){\cal D}(q^2),
\end{align}
where, using \3eqs{antighost-rel}{d0}{D},
\begin{align}
	\widehat{\alpha}(q^2)&=\alpha_0\frac{D(q^2)}{{\cal D}(q^2)}\left[\frac{F(q^2;\mu^2)/F(0;\mu^2)}{1-L(q^2;\mu^2)F(q^2;\mu^2)}\right]^2\\
	&\hspace{-.23cm}\overset{q^2\lesssim\mu^2}{=}\alpha_0\left[\frac{F(q^2;\mu^2)/F(0;\mu^2)}{1-L(q^2;\mu^2)F(q^2;\mu^2)}\right]^2.	
	\label{QCD-effchrg}
\end{align}
I have thus arrived at a definition of an effective charge $\widehat\alpha$ which is: ({\it i}) RG invariant; ({\it ii}) process independent; ({\it iii}) equivalent to the standard QCD running coupling in the UV; ({\it iv}) saturating at the IR to $\alpha_0$; and, last but not least, ({\it v}) parameter free, being expressed in terms of functions that can be computed using continuum and/or lattice methods. 

\begin{figure}[!t]
	\centering
	\includegraphics[scale=0.55]{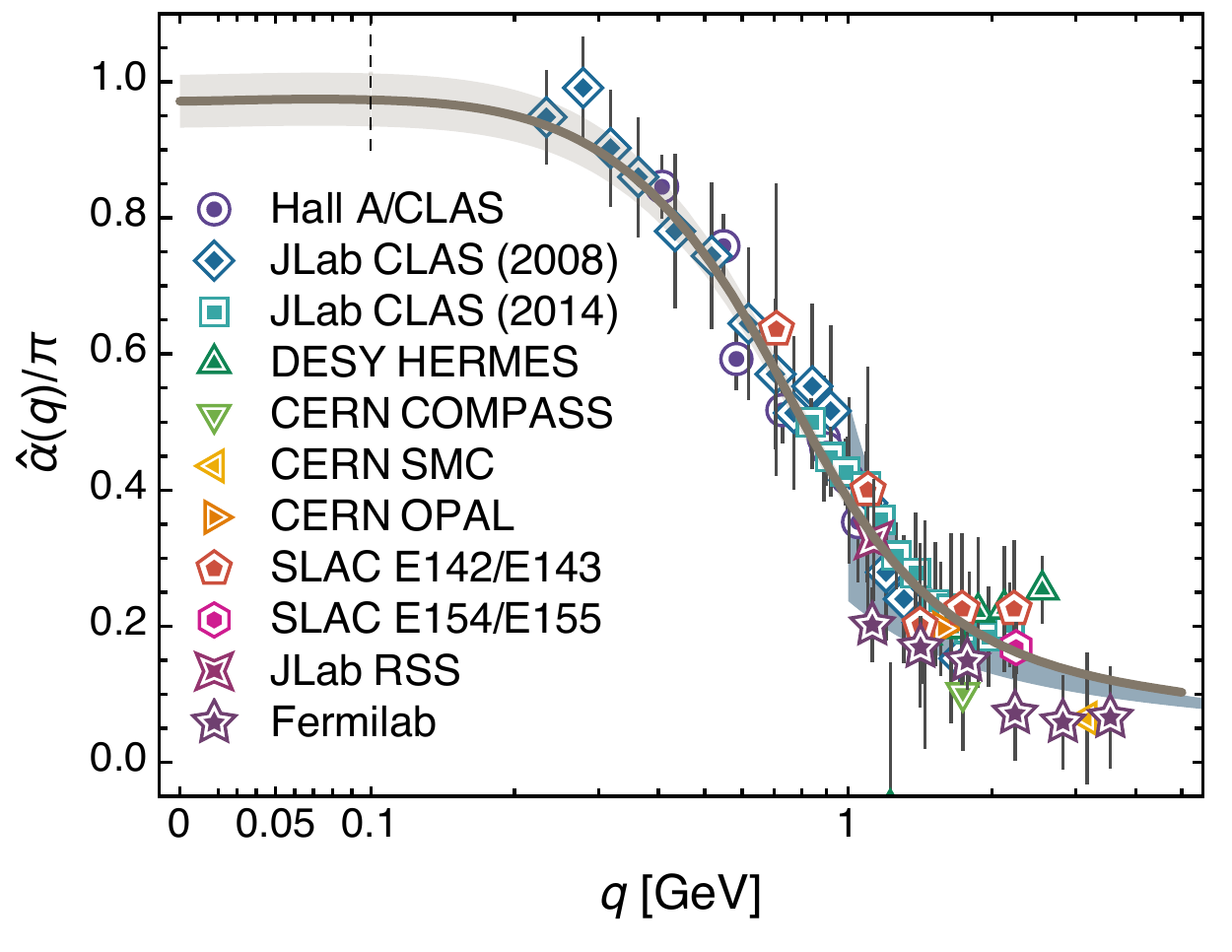}
	\caption{\label{fig:eff-chrg}The predicted QCD all-order effective charge obtained using~\1eq{QCD-effchrg} and the most precise unquenched lattice results for QCD's gauge sector~\cite{Cui:2019dwv}. Plotted are also world data on the process-dependent $\alpha_{g_1}$ defined via the Bjorken sum rule, see~\1eq{alpha-g1}.}
\end{figure}

Using lattice data for the gluon and ghost propagators obtained from gauge configurations with $N_f=3$ active (domain-wall) light quarks and a physical pion mass ($m_\pi=0.139$ GeV), and choosing a renormalization point $\mu=3.6$ GeV that lies within both the domain of reliable lattice output and perturbative QCD validity ($\Lambda_{\s{\mathrm{QCD}}}=0.58$ GeV with three active quark flavours), one obtains~\cite{Zafeiropoulos:2019flq,Cui:2019dwv}
\begin{align}
	g^2(\mu^2)&=4.44;& m_{\mathrm{gl}}(0;\mu^2)&=0.445\ \mathrm{GeV};&
	m_0&=0.428\ \mathrm{GeV},
\end{align}
which shows that the gluon mass is half that of the proton, $m_0\approx m_p/2$; and, finally,
\begin{align}
	\widehat{d}(0)&=16.6(4)\ \mathrm{GeV^{-2}};&
	\alpha_0=m_0^2\widehat{d}(0)=0.97(4)\pi.
\end{align}
The resulting effective charge is then represented by the continuous line in~\fig{fig:eff-chrg}, where I also plot for comparison the process-dependent effective charge\footnote{In a process-dependent approach one constructs an effective running coupling by using the leading-order term in the perturbative expansion of a given observable in terms of the canonical running coupling~\cite{Grunberg:1982fw}. Notice that process-dependence represents, however, a challenge as it hampers the ability of such a charge to predict any other observable.} $\alpha_{g_1}$ defined via the Bjorken sum rule~\cite{Bjorken:1966jh,Bjorken:1969mm}:
\begin{align}
	\int_0^1\!\mathrm{d}x\left[g_1^p(x,q^2)-g_1^n(x,q^2)\right]=\frac{g_A}6\left[1-\frac{\alpha_{g_1}(q^2)}{\pi}\right],
	\label{alpha-g1}
\end{align}
where: $g^{p,n}_1$ are the spin-dependent proton and neutron structure functions extracted from measurements using polarized targets; and $g_A$ is the nucleon flavour-singlet axial-charge. As can be clearly seen, there is an almost perfect match between the effective charges defined in \2eqs{QCD-effchrg}{alpha-g1}. While, on the one hand, the equivalence on the UV domain is guaranteed for any two reasonable definitions of QCD's effective charge (but notice that a detailed comparison shows that  sub-leading terms differ by just 4\%), on the other hand, the excellent match  below the scale at which perturbation theory would locate the Landau pole, is highly non-trivial, being a result of the careful inclusion of the gluon-ghost dynamics analyzed in the previous subsections. Summarizing, the identity between $\widehat{\alpha}$ and $\alpha_{g_1}$ singles out the Bjorken sum rule as a near direct means by which one can gain empirical insight into a QCD analogue of the Gell-Mann--Low effective charge~\cite{GellMann:1954fq}. 
 
Let me end by providing a continuous interpolation for $\widehat{\alpha}$ that can be used for modelling the PI charge in phenomenological studies. This can be achieved by setting
\begin{align}
	\widehat{\alpha}(q^2)&=\frac{\gamma_m\pi}{\log\frac{{\cal K}^2(q^2)}{\Lambda^2_{\s{\mathrm{QCD}}}}};&
	{\cal K}^2(q^2)=\frac{a^2_0+a_1q^2+q^4}{b_0+q^2},
	\label{nptcoupl}
\end{align}
where $\gamma_m=4(11-2/3 N_f)$. As the IR fixed point is renormalization group invariant, the interpolation coefficients $a_0$, $a_1$ and $b_0$ can be  then determined as a function of the (renormalization scheme dependent) parameter $\Lambda^2_{\s{\mathrm{QCD}}}$, so that, after they have been calculated within a reference scheme, they can be used to provide the PI coupling matching the perturbative tail in any other scheme. Clearly, the equality  between $\widehat{\alpha}$ and $\alpha_{g_1}$ obtained within the momentum subtraction scheme provides a natural candidate for the reference scheme in which the interpolation parameters can be calculated; and choosing\footnote{For phenomenological reasons this parametrization uses also the charm quark, so effectively the number of active quarks is 3 above the $s$ threshold, and 4 above the $c$ threshold (1.27 GeV). As shown in Fig.~3 of~\cite{Zafeiropoulos:2019flq}: there is no effect below the $c$ threshold (which has to be expected as the lack of a tree-level ghost-quark coupling implies that ghost related functions such as $F$ and $G$ are remarkably insensitive to the number of active fermions), and a marginal 5\% effect up to 3.5 GeV. The main difference is in the perturbative tail, that slows down when the number of flavours is increased.} $N_f=4$ so that $\Lambda=0.52$ GeV, a fitting to the numerical curve shown in Fig.~\ref{fig:eff-chrg} yields (all in GeV$^2$)
\begin{align}
	a_0&=0.5138;&
	a_1&=0.4814&
	b_0&=0.5952,
\end{align}
so that in a general renormalization scheme one has  
\begin{align}
	a_0&=1.9\Lambda^2_{\s{\mathrm{QCD}}};&
	a_1&=1.7805\Lambda^2_{\s{\mathrm{QCD}}};&
	b_0&=2.2010\Lambda^2_{\s{\mathrm{QCD}}}. 
\end{align}
Evidently, QCD's non-perturbative effects serve to replace the perturbative $q^2/\Lambda^2_{\s{\mathrm{QCD}}}$ in the argument of the $\log$ with the kernel ${\cal K}^2/\Lambda^2_{\s{\mathrm{QCD}}}$; and, as a result, the Landau pole at $q^2=\Lambda^2_{\s{\mathrm{QCD}}}$ will be replaced by the hadronic scale
\begin{align}
	\zeta_H=\left.{\cal K}(q^2)\right\vert_{q^2=\Lambda^2_{\s{\mathrm{QCD}}}}\approx1.413\Lambda_{\s{\mathrm{QCD}}}.
	\label{Hscale}
\end{align} 
The latter draws a natural border between soft and hard physics: the running coupling changes character at $\zeta_H$ bending towards its $\alpha_0$ saturation value. Modes with $q^2\lesssim\zeta_H^2$ are then screened and the theory is driven to a conformal phase.

\section{\label{sect.quarks}Quarks 2-point sector}

The last 2-point DSE I am going to study is the one corresponding to the quark propagator. For the quark self-energy $\Sigma$ one has then the following expression 
\begin{align}
	\Sigma(p)&=-C_fg^2\!\int_k\Delta^{\mu\nu}(q)\gamma_\mu S(k)\Gamma_\nu(k,q,-p),
	\label{sigma}
\end{align}
where: $C_{f}$ is the Casimir eigenvalue of the fundamental representation of the SU$(N)$ gauge group ($C_{f}=N/2-1/2N$); and 
\begin{align}
	i\Gamma^a_\mu(k,q,p)&=igt^a\Gamma_\mu(k,q,p);& \Gamma_\mu^{(0)}(k,q,p)=\gamma_\mu,
\end{align}
is the gluon-quark vertex (as usual, I assume all momenta entering into the vertex). In terms of the effective charge defined in Subsect.~\ref{ssec:eff-coupl}, the equation above becomes
\begin{align}
	\Sigma(p)=-4\pi C_f\!\int_k\widehat{\alpha}(q^2){\cal D}^{\mu\nu}(q)\gamma_\mu S(k)\Gamma_\nu(k,q,-p),
\end{align}
where I have defined ${\cal D}^{\mu\nu}={\cal D}P^{\mu\nu}$. The rewriting above thus bridges the study of QCD's gauge sector for determining, through an {\it ab-initio} computation, the theory's effective interaction (in what is referred as a top-down approach), with that body of work aimed at inferring that same interaction by fitting data using the DSE/BSE relevant to bound-state properties (in a bottom-up fashion)~\cite{Binosi:2014aea}.
 
Writing
\begin{align}
	S^{-1}(p;\mu)&=i\gamma{\cdot}p A(p^2;\mu^2)+B(p^2;\mu^2)\nonumber \\
	&=Z_2(\mu^2)(i\gamma{\cdot}k+m_{\mathrm{bare}})+Z_1(\mu^2)i\Sigma(p;\mu^2),
	\label{qgap}
\end{align} 
where $m_{\mathrm{bare}}$ is the bare current-quark mass appearing in the Lagrangian~\noeq{Linv}, and $Z_1$ (respectively, $Z_2$) represents the gluon-quark vertex (respectively, the quark wave-function) renormalization constant, the form factors $A$ and $B$ can be then obtained through suitable Dirac traces of the right-hand side. 

\begin{figure}[!t]
\centering
\includegraphics[scale=0.55]{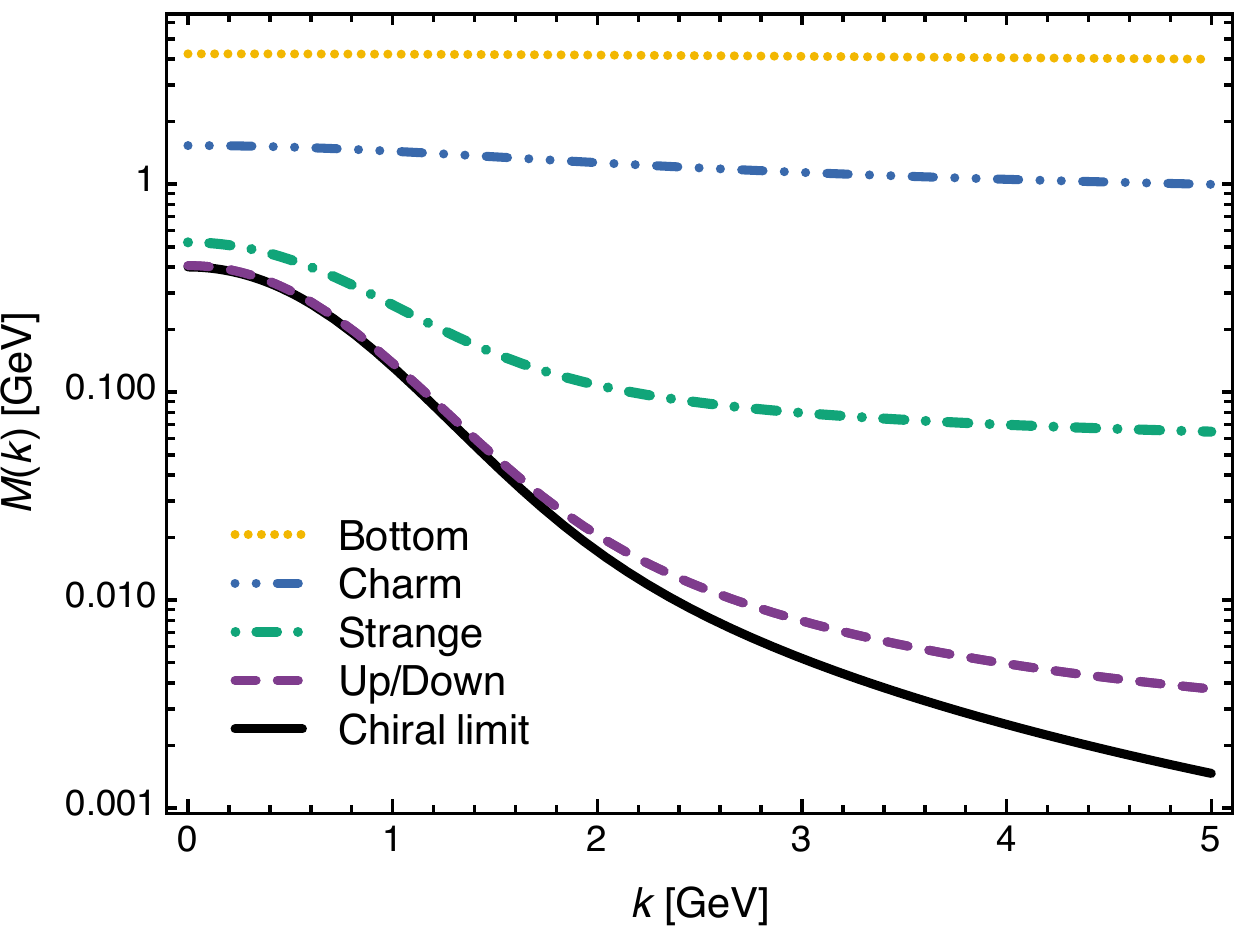}
\caption{\label{fig:gapmass} Dressed-quark mass function, $M$, at the renormalization point $\mu = 2$ GeV, obtained as the nonperturbative solution of the quark gap DSE~\noeq{qgap} using a vertex belonging to the class $\mathbb{G}_4$ defined in~Eq.~(11) of~\cite{Binosi:2016wcx}. At $p^2=0$, the chiral limit result is $M_0=0.40$ GeV; for light quarks $M_{u/d} =0.406\,$GeV; for the strange quark $M_{s} =0.526\,$GeV; and, finally, for the heavier quarks, $M_{c} =1.27$ GeV; $M_{b}=4.18$ GeV.}
\end{figure}

The most widely used truncation method to solve the DSE~\noeq{qgap} was introduced around twenty-five years ago~\cite{Munczek:1994zz,Bender:1996bb}; its leading-order term is the rainbow-ladder (RL) truncation, in which the full gluon-quark vertex appearing in~\1eq{sigma} is replaced by its tree-level value: $\Gamma_\nu=\Gamma_\nu^{(0)}=\gamma_\nu$. A solution of the DSE equation~\noeq{qgap} for the dressed-quark mass function $M=B/A$ employing however a beyond RL vertex~\cite{Binosi:2016wcx} is shown in~\fig{fig:gapmass}. Clearly, in the chiral limit (that is in the absence of Higgs couplings into QCD) the mass function displayed is fully non-perturbative, as no finite sum of perturbative diagrams can produce $M_{0}\neq 0$; and indeed continuum and lattice QCD~\cite{Bowman:2005vx} agree that massless partonic quarks acquire a momentum dependent mass function which is large at IR momenta.  This is dynamical chiral symmetry breaking (DCSB), a corollary of EHM: UV massless quarks acquire a large IR mass through interactions with their own gluon field. Notice that at $p^2=0$, $M_0 \approx 0.4\,$GeV, which is a typical scale for the constituent quark mass used in phenomenologically successful quark models \cite{Giannini:2015zia,Plessas:2015mpa,Eichmann:2016yit,Qin:2020rad}. With Higgs couplings reintroduced, the mass function becomes flavour dependent and its $p^2=0$ value is roughly the sum of $M_0$ and the appropriate current-quark mass.

The need to go beyond the RL truncation is because it provides a good description of hadronic bound-states in pseudoscalar and vector channels (when ignoring the non-Abelian anomaly is possible), as corrections to this truncation interfere destructively. However, when considering, {\it e.g.}, axial vector mesons the RL truncation is a poor approximation as \cite{Chang:2009zb,Chang:2010hb,Chang:2011ei} ({\it i}) some of the corrections starts interfering constructively and ({\it ii}) DCSB introduces novel contributions to Bethe-Salpeter kernels that enhance spin-orbit repulsion effects, thus magnifying even more the importance of those contributions in ({\it i}). Consequently, more sophisticated truncations are needed to describe axial-vector, scalar and tensor mesons.  In this connection, it is insufficient to improve upon RL truncation term-by-term because DCSB is essentially nonperturbative; thus, its contributions to Bethe-Salpeter kernels are missed in such a construction.  Alternatives have been developed \cite{Williams:2015cvx,Chang:2011ei,Qin:2016fwx,Qin:2020jig} and are being exploited in connection with light-quark mesons.  Similarly, the $\eta$-$\eta^\prime$ complex requires an essentially nonperturbative improvement of RL truncation \cite{Ding:2018xwy}.

\begin{figure}[!t]
	\centering
	\includegraphics[scale=0.3]{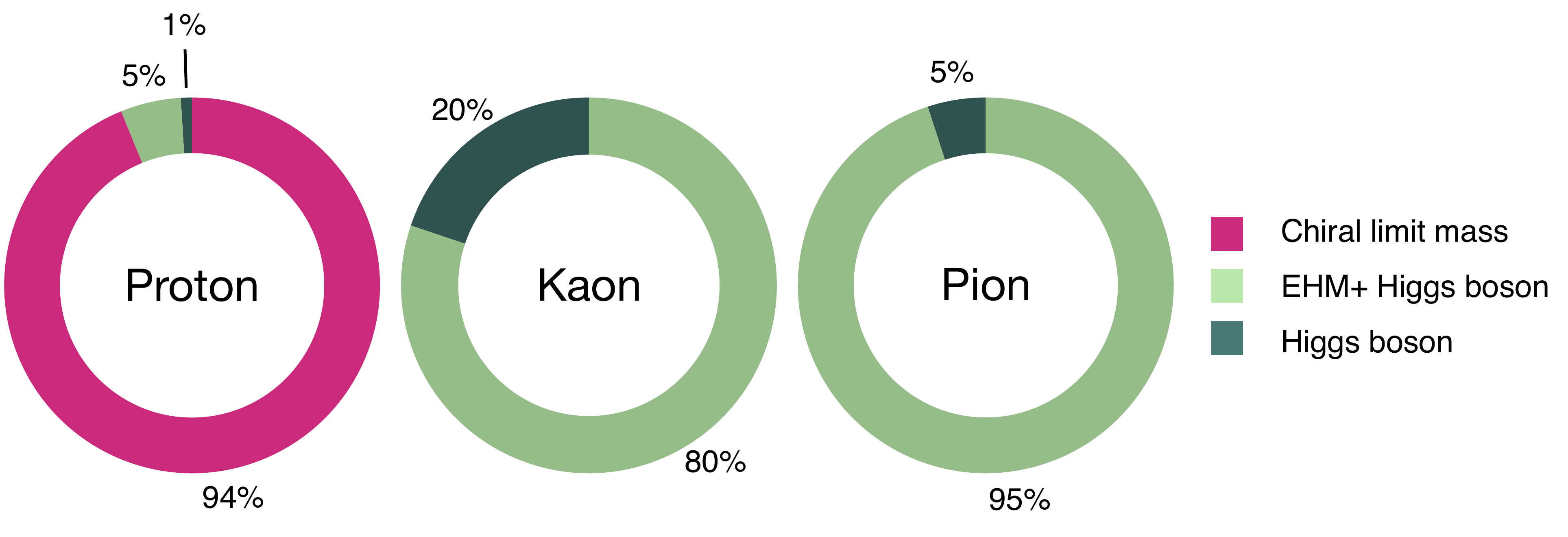}
	\caption{\label{fig:Mass-budgets} Proton, pion and kaon mass budgets at a $\zeta=2$ GeV scale. Owing to EHM, the proton's mass is large in the chiral limit; conversely and yet still owing to EHM via its dynamical chiral symmetry breaking corollary, the kaon and pion are massless in the absence of quark couplings to the Higgs boson.}
\end{figure}

\section{Phenomenology}

Once reformulated in the PT-BFM framework, the DSEs describing QCD's 2-point sector give us a qualitative and quantitative understanding of the simplest expressions of EHM: ({\it i}) the dynamical generation of a running gluon mass, leading to ({\it ii}) the acquisition of a running mass by the quarks (which would be massless in the absence of a Higgs mechanism) and therefore to ({\it iii}) DCSB with the emergence of its associated pseudoscalar Nambu-Goldstone bosons, the pions. In the following we will go further and try to elucidate what kind of effects does EHM imprint in QCD observables. 

\begin{figure}[!t]
	\centering
	\includegraphics[scale=0.55]{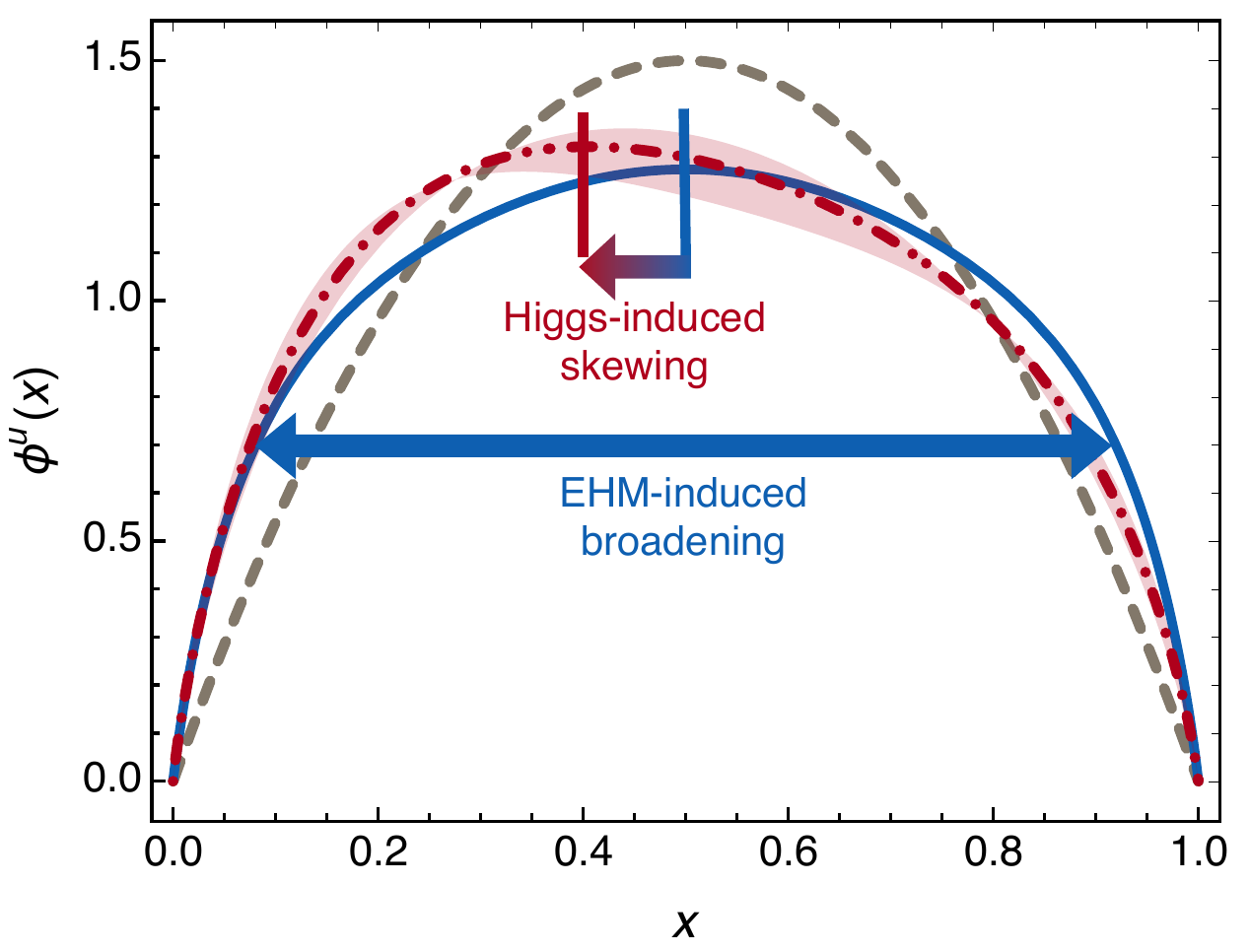}
	\caption{\label{fig:DAs-pion-kaon}DAs of the u quark in the pion (continuous, blue); the kaon (dot-dashed, red), and the asymptotic DA (dashed, brown), which is approached by all ground-state meson DAs as the ratio of the hadron mass over the energy scale associated with a given experiment goes to zero. As can be seen, at the scales accessible in  experiments, meson DAs display EHM-induced broadening; and Higgs-induced skewing (with the peak shifted away from $x = 0.5$) in systems defined by valence-quarks with different current-masses.}
\end{figure}

To begin with, EHM explains the general spectroscopic success of the constituent-quark picture.  Hadron masses are a global (volume-integrated) property: therefore, when using bound-state methods their values are by and large determined from the IR magnitude of the mass function of the hadron's defining valence quarks. Evidently, the volume integration focuses resolution on IR properties of the quasiparticle constituents\footnote{To be sure, even a carefully formulated momentum-independent interaction produces a decent description of hadron spectra \cite{Yin:2019bxe,Gutierrez-Guerrero:2019uwa}.}. The necessary IR scales are generated by the effective charge in~\fig{fig:eff-chrg}, which ultimately yields the mass functions in \fig{fig:gapmass}. Owing to EHM, the proton's mass is large in the chiral limit, and constitutes 94\% of its value at a scale of 2 GeV, see~\fig{fig:Mass-budgets}. However, the pion and kaon mass budgets are completely different: they are still due to EHM but this time through the realization of its DCSB corollary. These mesons are massless in the chiral limit, representing the Standard Model's Nambu-Goldstone modes; and for the pion the interference between the EHM and Higgs boson mass generation mechanism amounts to a staggering 95\% of its mass. The kaon lies in between these two extremes with the sum of valence-quark and -antiquark current-masses accounting for 20\% of its measured mass (4 times more than in the pion case). As such the pion and kaon provide an extraordinarily clear window onto understanding EHM and its modulation by Higgs-boson interactions~\cite{Arrington:2021biu,Aguilar:2019teb}.

This can be seen when studying the light-cone projection of a hadron wave-function leading to Distribution Amplitudes (DAs) and Functions (DFs) which measures the probability that a given parton carries a fraction $x$ of the hadron's total light-front momentum. Such distributions are in fact best evaluated at the hadronic scale $\zeta_H$~\noeq{Hscale} uniquely  determined from the process independent charge $\widehat{\alpha}$. At this scale the dressed quasiparticles obtained as solutions of the quark gap equation express all properties of the bound-state under scrutiny: for example, they carry all the hadron's momentum\footnote{Stated differently, at the hadronic scale $\zeta_H$, the nonperturbative quantum field equations of motion resum all bare-gluon and -quark contributions into the compound quasiparticle degrees-of-freedom in terms of which one chooses to resolve the problem; whereas, in a treatment of structure functions using a parton-basis Fock-space expansion, these contributions must  all be kept explicitly.}. At this scale, the pion and kaon DAs, show two hard-to-miss features, see~\fig{fig:DAs-pion-kaon}. First, they are very different from the asymptotic form that all such distributions are bound to approach as the ratio of the hadron mass over the energy scale associated with a given experiment goes to zero (brown dashed curve in~\fig{fig:DAs-pion-kaon}); the observed broadening is a direct consequence of EHM. Second, the Higgs-generated disparity in size between the strange-quark and the light up/down quarks current masses, which is roughly a factor of 25, is manifested as a small 20\% shift in the peak location of the kaon DA, see again~\fig{fig:DAs-pion-kaon}. In fact, this effect is controlled by the ratio of the kaon and pion decay constants ($f_K/f_\pi\approx1.2$) rather than the ratio of the strange and up/down quarks current masses.

\begin{figure}[!t]
	\centering
	\includegraphics[scale=0.55]{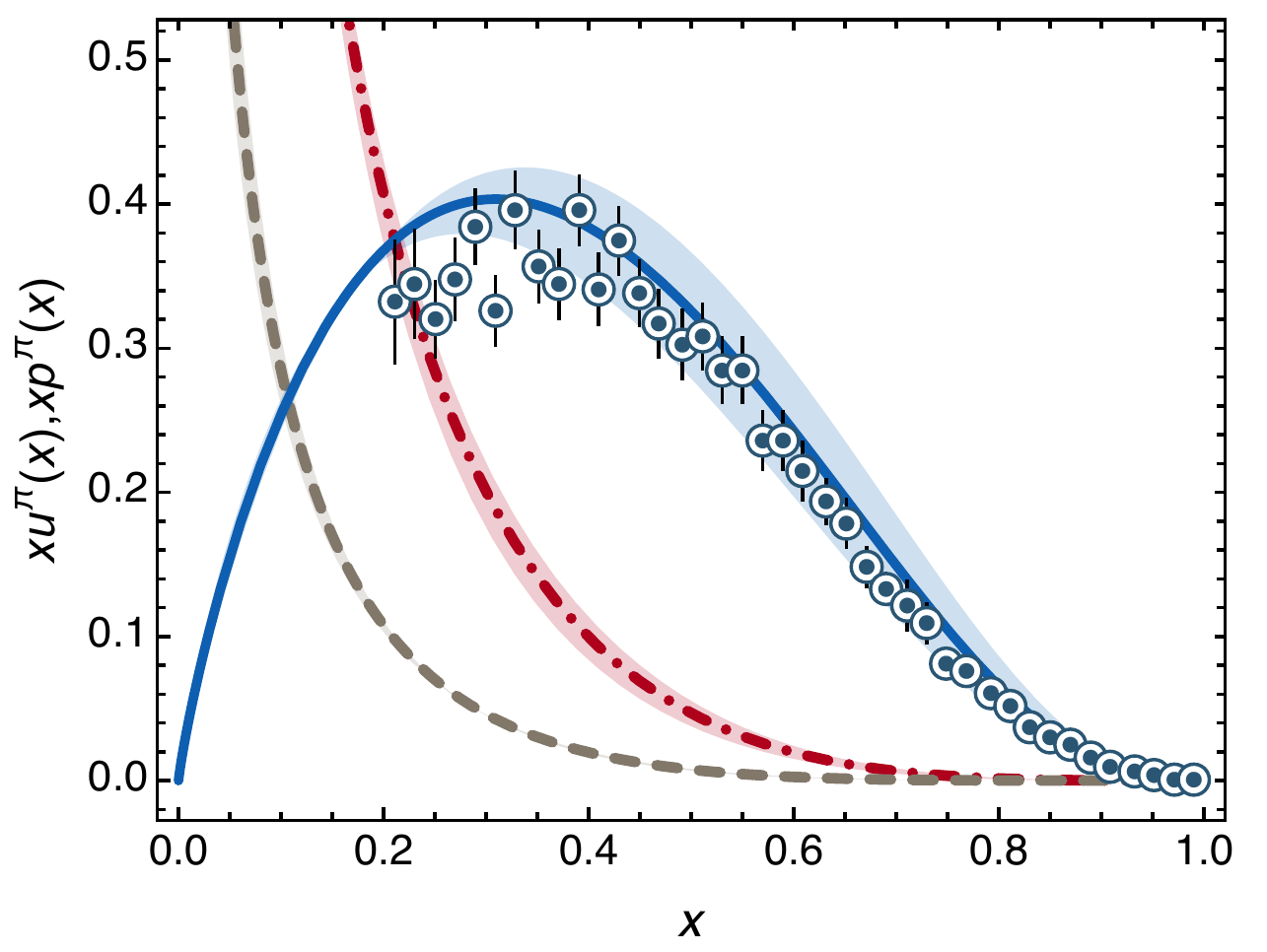}
	\caption{\label{fig:pionDF} Valence-quark (continuous, blue), gluon (dot-dashed, red) and sea-quark (dashed, brown) distribution functions evolved to $\zeta_5$. The data shown are those reported in~\cite{Conway:1989fs}, after rescaling according to the analysis of~\cite{Aicher:2010cb}.}
\end{figure}

At the hadronic scale $\zeta_H$ only valence quarks contribute to  pion and kaon DFs (which are accurately\footnote{Here by ``accurate'' I mean that computed values for physical quantities differ by less than any realistic estimate of input-model uncertainty.} obtained from the square of the corresponding DA): gluon and sea distributions are zero, and generated through (all-order) DGLAP evolution using the non-perturbative coupling~\noeq{nptcoupl}. This has to be contrasted with standard calculations in which $\zeta_H$ is a parameter and one starts with non-zero DFs for all particles. In the pion case, the scale relevant to the E615 experimental data for the $u$-quark DF within the pion is $\zeta_5=5.2$ GeV; and when evolved to this scale one obtains the comparison displayed in~\fig{fig:pionDF}~\cite{Cui:2020dlm,Cui:2020tdf}. The valence quarks carry a fraction of the pion's momentum given by the first Mellin moment of the DF; at $\zeta_5$ I find
\begin{align}
	\langle 2 xu^\pi(x)\rangle=\int_0^1\!\mathrm{d}x\,2xu^\pi(x)=0.41(4),
\end{align}
that is, at the scale relevant for the E615 experiment, 60\% of the pion's momentum is carried by gluons and sea-quarks. In addition, our result shows good agreement (with a $\chi^2/\mathrm{d.o.f}=1.66$) to the Next-to-Leading-Order re-analysis~\cite{Aicher:2010cb} of the original E615 experimental Drell-Yan data~\cite{Conway:1989fs}. Indeed, the EHM continuum approach illustrated here predicts a quadratic behavior at the end-points when $\zeta=\zeta_H$: $u(x)\sim(1-x)^2$. Hence one gets a power greater than 2 at the end-points at any experimentally accessible scale.
\begin{figure}
	\centering
	\includegraphics[scale=0.55]{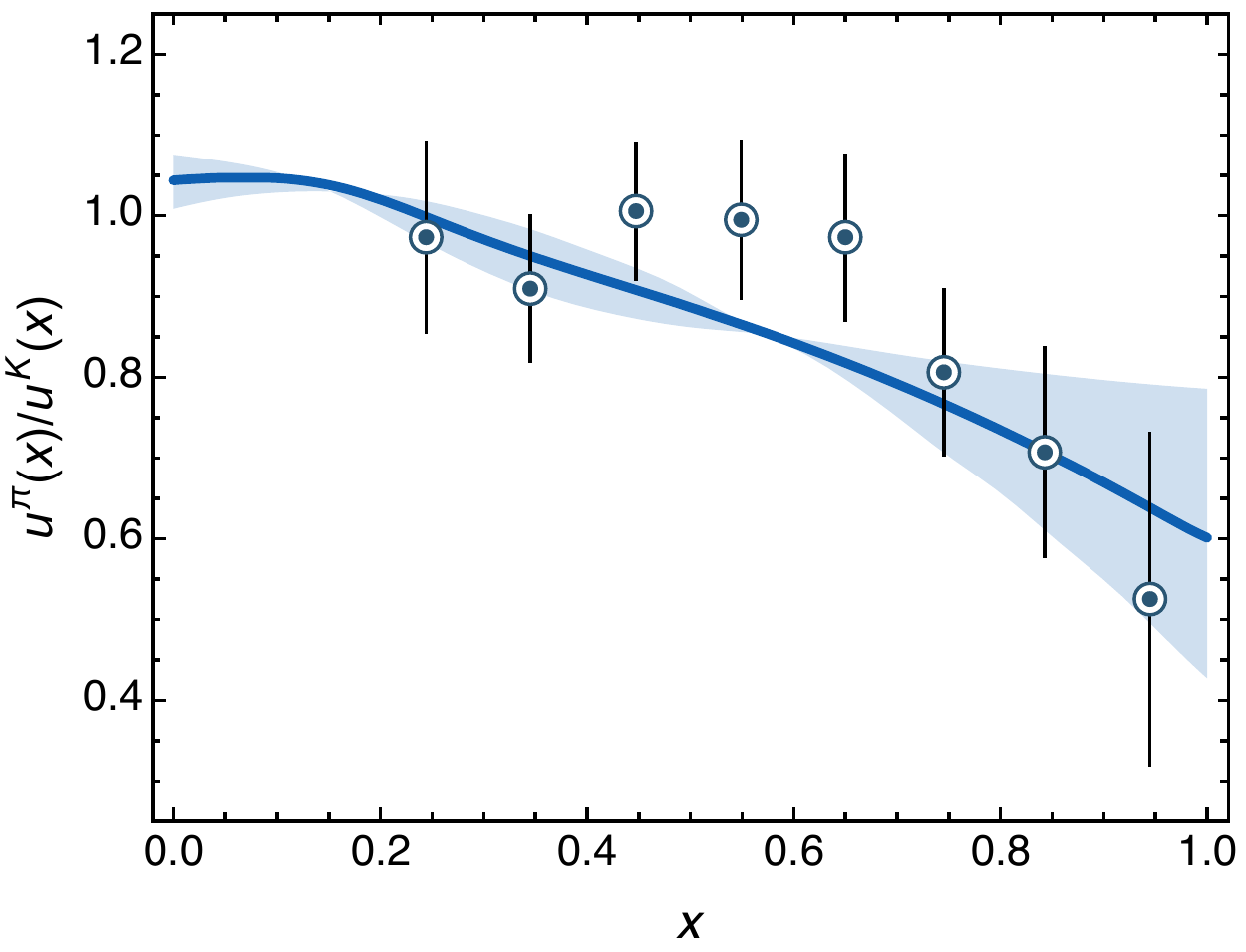}
	\caption{\label{fig:K-p-ratio}A comparison between the theoretically predicted~\cite{Cui:2020dlm,Cui:2020tdf} and experimentally measured~\cite{Saclay-CERN-CollegedeFrance-EcolePoly-Orsay:1980fhh} ratio of the $u$-quark distribution functions in the kaon and in the pion.}
\end{figure}

For the kaon there are no experimental data for the DF; however, the NA3 experiment has measured the ratio of the $u$ quark DFs in the kaon and the pion~\cite{Saclay-CERN-CollegedeFrance-EcolePoly-Orsay:1980fhh}. The results are shown in~\fig{fig:K-p-ratio} together with the predicted ratio of~\cite{Cui:2020dlm,Cui:2020tdf}, which shows a fair agreement with them. It should be noticed however that this ratio is not very sensitive to the details of the pion and kaon DFs, so that it is not a good discriminator between pictures of meson structure.    

Novel developments in the theory and phenomenology of DFs, related to the effective charge concept described here, are detailed in~\cite{Cui:2021mom,Cui:2022bxn}; and the extension of these ideas to proton and neutron DFs is presented in~\cite{Chang:2022jri}. 

\section{Conclusions}

There is currently an intense experimental program being pursued worldwide in hadro-particle physics with the operation, construction and planning of many facilities: Jefferson Lab running at 12 GeV; AMBER at CERN; and the electron ion colliders planned in the USA (EIC) and China (EicC). These experimental efforts, coupled with our increased ability to treat QCD's IR physics through {\it ab-initio} methods based on solving the theory's equations of motion (DSEs and BSEs {\it in primis}) and the associated appearance of Emergent Hadron Mass along the lines described in these notes, are expected to contribute to the full understanding of QCD; finally, almost fifty years after its formulation.

\begin{acknowledgements}
These notes encompass twenty years of research on the subject carried out in collaboration with many people, to whom I am greatly indebted.  
\end{acknowledgements}

\newpage




\begin{thebibliography}{100}
\providecommand{\url}[1]{{#1}}
\providecommand{\urlprefix}{URL }
\expandafter\ifx\csname urlstyle\endcsname\relax
  \providecommand{\doi}[1]{DOI \discretionary{}{}{}#1}\else
  \providecommand{\doi}{DOI \discretionary{}{}{}\begingroup
  \urlstyle{rm}\Url}\fi

\bibitem{Gross:1973id}
D.J. Gross, F.~Wilczek, Phys. Rev. Lett. \textbf{30}, 1343 (1973).
\newblock \doi{10.1103/PhysRevLett.30.1343}

\bibitem{Politzer:1973fx}
H.D. Politzer, Phys. Rev. Lett. \textbf{30}, 1346 (1973)

\bibitem{ATLAS:2012yve}
G.~Aad, et~al., Phys. Lett. B \textbf{716}, 1 (2012).
\newblock \doi{10.1016/j.physletb.2012.08.020}

\bibitem{CMS:2012qbp}
S.~Chatrchyan, et~al., Phys. Lett. B \textbf{716}, 30 (2012).
\newblock \doi{10.1016/j.physletb.2012.08.021}

\bibitem{Roberts:2020hiw}
C.D. Roberts, Symmetry \textbf{12}(9), 1468 (2020).
\newblock \doi{10.3390/sym12091468}

\bibitem{Nakanishi:1966cq}
N.~Nakanishi, Prog. Theor. Phys. \textbf{35}(6), 1111 (1966)

\bibitem{Lautrup:1966cq}
B.~Lautrup, Mat. Fys. Medd. Dan. Vid. Selsk. \textbf{35}(11), 1 (1966)

\bibitem{Becchi:1974md}
C.~Becchi, A.~Rouet, R.~Stora, Commun. Math. Phys. \textbf{42}, 127 (1975)

\bibitem{Becchi:1975nq}
C.~Becchi, A.~Rouet, R.~Stora, Annals Phys. \textbf{98}, 287 (1976)

\bibitem{Tyutin:1975qk}
I.V. Tyutin, LEBEDEV-75-39, arXiv:0812.0580 [hep-th] (1975)

\bibitem{Fujikawa:1972fe}
K.~Fujikawa, B.W. Lee, A.I. Sanda, Phys. Rev. \textbf{D6}, 2923 (1972)

\bibitem{Binosi:2009qm}
D.~Binosi, J.~Papavassiliou, Phys. Rept. \textbf{479}, 1 (2009).
\newblock \doi{10.1016/j.physrep.2009.05.001}

\bibitem{Roberts:1994dr}
C.D. Roberts, A.G. Williams, Prog. Part. Nucl. Phys. \textbf{33}, 477 (1994)

\bibitem{Aguilar:2021okw}
A.C. Aguilar, C.O. Ambr\'osio, F.~De~Soto, M.N. Ferreira, B.M. Oliveira,
  J.~Papavassiliou, J.~Rodr\'\i{}guez-Quintero, Phys. Rev. D \textbf{104}(5),
  054028 (2021).
\newblock \doi{10.1103/PhysRevD.104.054028}

\bibitem{Schwinger:1962tn}
J.S. Schwinger, Phys. Rev. \textbf{125}, 397 (1962)

\bibitem{Schwinger:1962tp}
J.S. Schwinger, Phys. Rev. \textbf{128}, 2425 (1962)

\bibitem{Cornwall:1981zr}
J.M. Cornwall, Phys. Rev. \textbf{D26}, 1453 (1982)

\bibitem{Aguilar:2008xm}
A.C. Aguilar, D.~Binosi, J.~Papavassiliou, Phys. Rev. \textbf{D78}, 025010
  (2008).
\newblock \doi{10.1103/PhysRevD.78.025010}

\bibitem{Cucchieri:2007md}
A.~Cucchieri, T.~Mendes, PoS \textbf{LAT2007}, 297 (2007)

\bibitem{Cucchieri:2007rg}
A.~Cucchieri, T.~Mendes, Phys.~Rev.~Lett. \textbf{100}, 241601 (2008).
\newblock \doi{10.1103/PhysRevLett.100.241601}

\bibitem{Bowman:2007du}
P.O. Bowman, et~al., Phys. Rev. \textbf{D76}, 094505 (2007)

\bibitem{Bogolubsky:2007ud}
I.L. Bogolubsky, E.M. Ilgenfritz, M.~Muller-Preussker, A.~Sternbeck, PoS
  \textbf{LATTICE2007}, 290 (2007).
\newblock \doi{10.22323/1.042.0290}

\bibitem{Bogolubsky:2009dc}
I.~Bogolubsky, E.~Ilgenfritz, M.~Muller-Preussker, A.~Sternbeck, Phys. Lett.
  \textbf{B676}, 69 (2009).
\newblock \doi{10.1016/j.physletb.2009.04.076}

\bibitem{Oliveira:2009eh}
O.~Oliveira, P.~Silva, PoS \textbf{LAT2009}, 226 (2009)

\bibitem{Cucchieri:2009zt}
A.~Cucchieri, T.~Mendes, Phys.~Rev.~ \textbf{D81}, 016005 (2010).
\newblock \doi{10.1103/PhysRevD.81.016005}

\bibitem{Cucchieri:2010xr}
A.~Cucchieri, T.~Mendes, PoS \textbf{QCD-TNT09}, 026 (2009)

\bibitem{Ayala:2012pb}
A.~Ayala, A.~Bashir, D.~Binosi, M.~Cristoforetti, J.~Rodriguez-Quintero, Phys.
  Rev. \textbf{D86}, 074512 (2012).
\newblock \doi{10.1103/PhysRevD.86.074512}

\bibitem{Binosi:2016xxu}
D.~Binosi, C.D. Roberts, J.~Rodriguez-Quintero, Phys. Rev. D \textbf{95}(11),
  114009 (2017).
\newblock \doi{10.1103/PhysRevD.95.114009}

\bibitem{Bicudo:2015rma}
P.~Bicudo, D.~Binosi, N.~Cardoso, O.~Oliveira, P.J. Silva, Phys. Rev.
  \textbf{D92}(11), 114514 (2015).
\newblock \doi{10.1103/PhysRevD.92.114514}

\bibitem{Dudal:2008sp}
D.~Dudal, J.A. Gracey, S.P. Sorella, N.~Vandersickel, H.~Verschelde, Phys. Rev.
  \textbf{D78}, 065047 (2008).
\newblock \doi{10.1103/PhysRevD.78.065047}

\bibitem{Tissier:2011ey}
M.~Tissier, N.~Wschebor, Phys. Rev. \textbf{D84}, 045018 (2011).
\newblock \doi{10.1103/PhysRevD.84.045018}

\bibitem{Cyrol:2016tym}
A.K. Cyrol, L.~Fister, M.~Mitter, J.M. Pawlowski, N.~Strodthoff, Phys. Rev.
  \textbf{D94}(5), 054005 (2016).
\newblock \doi{10.1103/PhysRevD.94.054005}

\bibitem{Huber:2018ned}
M.Q. Huber, Phys. Rept. \textbf{879}, 1 (2020).
\newblock \doi{10.1016/j.physrep.2020.04.004}

\bibitem{Cornwall:1989gv}
J.M. Cornwall, J.~Papavassiliou, Phys. Rev. \textbf{D40}, 3474 (1989)

\bibitem{Binosi:2002ft}
D.~Binosi, J.~Papavassiliou, Phys. Rev. \textbf{D66}, 111901(R) (2002)

\bibitem{Binosi:2003rr}
D.~Binosi, J.~Papavassiliou, J.Phys.G \textbf{G30}, 203 (2004).
\newblock \doi{10.1088/0954-3899/30/2/017}

\bibitem{Binosi:2004qe}
D.~Binosi, J. Phys. \textbf{G30}, 1021 (2004)

\bibitem{Abbott:1980hw}
L.F. Abbott, Nucl. Phys. \textbf{B185}, 189 (1981)

\bibitem{Abbott:1981ke}
L.F. Abbott, Acta Phys. Polon. \textbf{B13}, 33 (1982)

\bibitem{Aguilar:2006gr}
A.C. Aguilar, J.~Papavassiliou, JHEP \textbf{12}, 012 (2006)

\bibitem{Binosi:2007pi}
D.~Binosi, J.~Papavassiliou, Phys.~Rev.~ \textbf{D77}, 061702 (2008).
\newblock \doi{10.1103/PhysRevD.77.061702}

\bibitem{Binosi:2008qk}
D.~Binosi, J.~Papavassiliou, JHEP \textbf{0811}, 063 (2008).
\newblock \doi{10.1088/1126-6708/2008/11/063}

\bibitem{Jackiw:1973tr}
R.~Jackiw, K.~Johnson, Phys. Rev. \textbf{D8}, 2386 (1973)

\bibitem{Jackiw:1973ha}
R.~Jackiw, In *Erice 1973, Proceedings, Laws Of Hadronic Matter*, New York
  1975, 225-251 and M I T Cambridge - COO-3069-190 (73,REC.AUG 74) 23p  (1973)

\bibitem{Cornwall:1973ts}
J.M. Cornwall, R.E. Norton, Phys. Rev. \textbf{D8}, 3338 (1973)

\bibitem{Eichten:1974et}
E.~Eichten, F.~Feinberg, Phys. Rev. \textbf{D10}, 3254 (1974)

\bibitem{Poggio:1974qs}
E.C. Poggio, E.~Tomboulis, S.H.H. Tye, Phys. Rev. \textbf{D11}, 2839 (1975).
\newblock \doi{10.1103/PhysRevD.11.2839}

\bibitem{Pilaftsis:1996fh}
A.~Pilaftsis, Nucl. Phys. \textbf{B487}, 467 (1997)

\bibitem{Papavassiliou:1999az}
J.~Papavassiliou, Phys. Rev. Lett. \textbf{84}, 2782 (2000)

\bibitem{Binosi:2013cea}
D.~Binosi, A.~Quadri, Phys.~Rev.~ \textbf{D88}, 085036 (2013).
\newblock \doi{10.1103/PhysRevD.88.085036}

\bibitem{Binosi:2002ez}
D.~Binosi, J.~Papavassiliou, Phys.~Rev.~ \textbf{D66}, 025024 (2002).
\newblock \doi{10.1103/PhysRevD.66.025024}

\bibitem{Grassi:1999tp}
P.A. Grassi, T.~Hurth, M.~Steinhauser, Annals Phys. \textbf{288}, 197 (2001)

\bibitem{Batalin:1977pb}
I.A. Batalin, G.A. Vilkovisky, Phys. Lett. \textbf{B69}, 309 (1977)

\bibitem{Batalin:1983ggl}
I.A. Batalin, G.A. Vilkovisky, Phys. Rev. D \textbf{28}, 2567 (1983).
\newblock \doi{10.1103/PhysRevD.28.2567}.
\newblock [Erratum: Phys.Rev.D 30, 508 (1984)]

\bibitem{Grassi:2001zz}
P.A. Grassi, T.~Hurth, M.~Steinhauser, Nucl. Phys. \textbf{B610}, 215 (2001)

\bibitem{Aguilar:2009nf}
A.C. Aguilar, D.~Binosi, J.~Papavassiliou, J.~Rodriguez-Quintero, Phys. Rev.
  \textbf{D80}, 085018 (2009).
\newblock \doi{10.1103/PhysRevD.80.085018}

\bibitem{Aguilar:2009pp}
A.~Aguilar, D.~Binosi, J.~Papavassiliou, JHEP \textbf{0911}, 066 (2009).
\newblock \doi{10.1088/1126-6708/2009/11/066}

\bibitem{Kugo:1979gm}
T.~Kugo, I.~Ojima, Prog. Theor. Phys. Suppl. \textbf{66}, 1 (1979)

\bibitem{Aguilar:2009ke}
A.C. Aguilar, J.~Papavassiliou, Phys.~Rev.~ \textbf{D81}, 034003 (2010).
\newblock \doi{10.1103/PhysRevD.81.034003}

\bibitem{Aguilar:2016vin}
A.C. Aguilar, D.~Binosi, C.T. Figueiredo, J.~Papavassiliou, Phys. Rev.
  \textbf{D94}(4), 045002 (2016).
\newblock \doi{10.1103/PhysRevD.94.045002}

\bibitem{Smit:1974je}
J.~Smit, Phys. Rev. \textbf{D10}, 2473 (1974).
\newblock \doi{10.1103/PhysRevD.10.2473}

\bibitem{Aguilar:2011xe}
A.~Aguilar, D.~Ibanez, V.~Mathieu, J.~Papavassiliou, Phys.~Rev.~ \textbf{D85},
  014018 (2012).
\newblock \doi{10.1103/PhysRevD.85.014018}

\bibitem{Binosi:2012sj}
D.~Binosi, D.~Iba\~nez, J.~Papavassiliou, Phys. Rev. \textbf{D86}, 085033
  (2012).
\newblock \doi{10.1103/PhysRevD.86.085033}

\bibitem{Aguilar:2016ock}
A.C. Aguilar, D.~Binosi, J.~Papavassiliou, Phys. Rev. \textbf{D95}(3), 034017
  (2017).
\newblock \doi{10.1103/PhysRevD.95.034017}

\bibitem{Binosi:2017rwj}
D.~Binosi, J.~Papavassiliou, Phys. Rev. D \textbf{97}(5), 054029 (2018).
\newblock \doi{10.1103/PhysRevD.97.054029}

\bibitem{Ball:1980ax}
J.S. Ball, T.W. Chiu, Phys. Rev. \textbf{D22}, 2550 (1980)

\bibitem{Alkofer:2008dt}
R.~Alkofer, M.Q. Huber, K.~Schwenzer, Eur. Phys. J. \textbf{C62}, 761 (2009).
\newblock \doi{10.1140/epjc/s10052-009-1066-3}

\bibitem{Pelaez:2013cpa}
M.~Pelaez, M.~Tissier, N.~Wschebor, Phys.~Rev.~ \textbf{D88}, 125003 (2013).
\newblock \doi{10.1103/PhysRevD.88.125003}

\bibitem{Aguilar:2013vaa}
A.C. Aguilar, D.~Binosi, D.~Iba{\~n}ez, J.~Papavassiliou, Phys. Rev.
  \textbf{D89}, 085008 (2014).
\newblock \doi{10.1103/PhysRevD.89.085008}

\bibitem{Blum:2014gna}
A.~Blum, M.Q. Huber, M.~Mitter, L.~von Smekal, Phys.~Rev.~ \textbf{D89}, 061703
  (2014).
\newblock \doi{10.1103/PhysRevD.89.061703}

\bibitem{Eichmann:2014xya}
G.~Eichmann, R.~Williams, R.~Alkofer, M.~Vujinovic, Phys.~Rev.~ \textbf{D89},
  105014 (2014).
\newblock \doi{10.1103/PhysRevD.89.105014}

\bibitem{Williams:2015cvx}
R.~Williams, C.S. Fischer, W.~Heupel, Phys. Rev. \textbf{D93}(3), 034026
  (2016).
\newblock \doi{10.1103/PhysRevD.93.034026}

\bibitem{Cucchieri:2006tf}
A.~Cucchieri, A.~Maas, T.~Mendes, Phys.~Rev.~ \textbf{D74}, 014503 (2006).
\newblock \doi{10.1103/PhysRevD.74.014503}

\bibitem{Cucchieri:2008qm}
A.~Cucchieri, A.~Maas, T.~Mendes, Phys.~Rev.~ \textbf{D77}, 094510 (2008).
\newblock \doi{10.1103/PhysRevD.77.094510}

\bibitem{Athenodorou:2016oyh}
A.~Athenodorou, D.~Binosi, P.~Boucaud, F.~De~Soto, J.~Papavassiliou,
  J.~Rodriguez-Quintero, S.~Zafeiropoulos, Phys. Lett. \textbf{B761}, 444
  (2016).
\newblock \doi{10.1016/j.physletb.2016.08.065}

\bibitem{Boucaud:2017obn}
P.~Boucaud, F.~De~Soto, J.~Rodr{\'\i}guez-Quintero, S.~Zafeiropoulos, Phys.
  Rev. \textbf{D95}(11), 114503 (2017).
\newblock \doi{10.1103/PhysRevD.95.114503}

\bibitem{Duarte:2016ieu}
A.G. Duarte, O.~Oliveira, P.J. Silva, Phys. Rev. \textbf{D94}(7), 074502
  (2016).
\newblock \doi{10.1103/PhysRevD.94.074502}

\bibitem{Aguilar:2021lke}
A.C. Aguilar, F.~De~Soto, M.N. Ferreira, J.~Papavassiliou,
  J.~Rodr\'\i{}guez-Quintero, Phys. Lett. B \textbf{818}, 136352 (2021).
\newblock \doi{10.1016/j.physletb.2021.136352}

\bibitem{Aguilar:2017dco}
A.C. Aguilar, D.~Binosi, C.T. Figueiredo, J.~Papavassiliou, Eur. Phys. J. C
  \textbf{78}(3), 181 (2018).
\newblock \doi{10.1140/epjc/s10052-018-5679-2}

\bibitem{Boucaud:2018xup}
P.~Boucaud, F.~De~Soto, K.~Raya, J.~Rodr\'\i{}guez-Quintero, S.~Zafeiropoulos,
  Phys. Rev. D \textbf{98}(11), 114515 (2018).
\newblock \doi{10.1103/PhysRevD.98.114515}

\bibitem{Sternbeck:2006rd}
A.~Sternbeck, hep-lat/0609016 (2006).

\bibitem{Cucchieri:2009xxr}
A.~Cucchieri, T.~Mendes, PoS \textbf{QCD-TNT09}, 026 (2009).
\newblock \doi{10.22323/1.087.0026}

\bibitem{Huber:2015ria}
M.Q. Huber, Phys.~Rev.~ \textbf{D91}(8), 085018 (2015).
\newblock \doi{10.1103/PhysRevD.91.085018}

\bibitem{Aguilar:2015nqa}
A.~Aguilar, D.~Binosi, J.~Papavassiliou, Phys.~Rev.~ \textbf{D91}(8), 085014
  (2015).
\newblock \doi{10.1103/PhysRevD.91.085014}

\bibitem{Binosi:2002vk}
D.~Binosi, J.~Papavassiliou, Nucl.Phys.Proc.Suppl. \textbf{121}, 281 (2003).
\newblock \doi{10.1016/S0920-5632(03)01862-0}

\bibitem{Binosi:2016nme}
D.~Binosi, C.~Mezrag, J.~Papavassiliou, C.D. Roberts, J.~Rodriguez-Quintero,
  Phys. Rev. \textbf{D96}(5), 054026 (2017).
\newblock \doi{10.1103/PhysRevD.96.054026}

\bibitem{Rodriguez-Quintero:2018wma}
J.~Rodr\'\i{}guez-Quintero, D.~Binosi, C.~Mezrag, J.~Papavassiliou, C.D.
  Roberts, Few Body Syst. \textbf{59}(6), 121 (2018).
\newblock \doi{10.1007/s00601-018-1437-0}

\bibitem{Cui:2019dwv}
Z.F. Cui, J.L. Zhang, D.~Binosi, F.~de~Soto, C.~Mezrag, J.~Papavassiliou, C.D.
  Roberts, J.~Rodr\'\i{}guez-Quintero, J.~Segovia, S.~Zafeiropoulos, Chin.
  Phys. C \textbf{44}(8), 083102 (2020).
\newblock \doi{10.1088/1674-1137/44/8/083102}

\bibitem{Zafeiropoulos:2019flq}
S.~Zafeiropoulos, P.~Boucaud, F.~De~Soto, J.~Rodr\'\i{}guez-Quintero,
  J.~Segovia, Phys. Rev. Lett. \textbf{122}(16), 162002 (2019).
\newblock \doi{10.1103/PhysRevLett.122.162002}

\bibitem{Grunberg:1982fw}
G.~Grunberg, Phys. Rev. \textbf{D29}, 2315 (1984)

\bibitem{Bjorken:1966jh}
J.D. Bjorken, Phys. Rev. \textbf{148}, 1467 (1966).
\newblock \doi{10.1103/PhysRev.148.1467}

\bibitem{Bjorken:1969mm}
J.D. Bjorken, Phys. Rev. D \textbf{1}, 1376 (1970).
\newblock \doi{10.1103/PhysRevD.1.1376}

\bibitem{GellMann:1954fq}
M.~Gell-Mann, F.E. Low, Phys. Rev. \textbf{95}, 1300 (1954)

\bibitem{Binosi:2014aea}
D.~Binosi, L.~Chang, J.~Papavassiliou, C.D. Roberts, Phys. Lett. \textbf{B742},
  183 (2015).
\newblock \doi{10.1016/j.physletb.2015.01.031}

\bibitem{Binosi:2016wcx}
D.~Binosi, L.~Chang, J.~Papavassiliou, S.X. Qin, C.D. Roberts, Phys. Rev. D
  \textbf{95}(3), 031501 (2017).
\newblock \doi{10.1103/PhysRevD.95.031501}

\bibitem{Munczek:1994zz}
H.~Munczek, Phys.~Rev.~ \textbf{D52}, 4736 (1995).
\newblock \doi{10.1103/PhysRevD.52.4736}

\bibitem{Bender:1996bb}
A.~Bender, C.D. Roberts, L.~Von~Smekal, Phys. Lett. B \textbf{380}, 7 (1996).
\newblock \doi{10.1016/0370-2693(96)00372-3}

\bibitem{Bowman:2005vx}
P.O. Bowman, U.M. Heller, D.B. Leinweber, M.B. Parappilly, A.G. Williams, J.b.
  Zhang, Phys. Rev. D \textbf{71}, 054507 (2005).
\newblock \doi{10.1103/PhysRevD.71.054507}

\bibitem{Giannini:2015zia}
M.M. Giannini, E.~Santopinto, Chin. J. Phys. \textbf{53}, 020301 (2015).
\newblock \doi{10.6122/CJP.20150120}

\bibitem{Plessas:2015mpa}
W.~Plessas, Int. J. Mod. Phys. A \textbf{30}(02), 1530013 (2015).
\newblock \doi{10.1142/S0217751X15300136}

\bibitem{Eichmann:2016yit}
G.~Eichmann, H.~Sanchis-Alepuz, R.~Williams, R.~Alkofer, C.S. Fischer, Prog.
  Part. Nucl. Phys. \textbf{91}, 1 (2016).
\newblock \doi{10.1016/j.ppnp.2016.07.001}

\bibitem{Qin:2020rad}
S.x. Qin, C.D. Roberts, Chin. Phys. Lett. \textbf{37}(12), 121201 (2020).
\newblock \doi{10.1088/0256-307X/37/12/121201}

\bibitem{Chang:2009zb}
L.~Chang, C.D. Roberts, Phys. Rev. Lett. \textbf{103}, 081601 (2009).
\newblock \doi{10.1103/PhysRevLett.103.081601}

\bibitem{Chang:2010hb}
L.~Chang, Y.X. Liu, C.D. Roberts, Phys.~Rev.~Lett. \textbf{106}, 072001 (2011).
\newblock \doi{10.1103/PhysRevLett.106.072001}

\bibitem{Chang:2011ei}
L.~Chang, C.D. Roberts, Phys. Rev. \textbf{C85}, 052201 (2012).
\newblock \doi{10.1103/PhysRevC.85.052201}

\bibitem{Qin:2016fwx}
S.x. Qin, Few Body Syst. \textbf{57}(11), 1059 (2016).
\newblock \doi{10.1007/s00601-016-1149-2}

\bibitem{Qin:2020jig}
S.X. Qin, C.D. Roberts, Chin. Phys. Lett. \textbf{38}(7), 071201 (2021).
\newblock \doi{10.1088/0256-307X/38/7/071201}

\bibitem{Ding:2018xwy}
M.~Ding, K.~Raya, A.~Bashir, D.~Binosi, L.~Chang, M.~Chen, C.D. Roberts, Phys.
  Rev. D \textbf{99}(1), 014014 (2019).
\newblock \doi{10.1103/PhysRevD.99.014014}

\bibitem{Yin:2019bxe}
P.L. Yin, C.~Chen, G.a. Krein, C.D. Roberts, J.~Segovia, S.S. Xu, Phys. Rev. D
  \textbf{100}(3), 034008 (2019).
\newblock \doi{10.1103/PhysRevD.100.034008}

\bibitem{Gutierrez-Guerrero:2019uwa}
L.X. Guti\'errez-Guerrero, A.~Bashir, M.A. Bedolla, E.~Santopinto, Phys. Rev. D
  \textbf{100}(11), 114032 (2019).
\newblock \doi{10.1103/PhysRevD.100.114032}

\bibitem{Arrington:2021biu}
J.~Arrington, et~al., J. Phys. G \textbf{48}(7), 075106 (2021).
\newblock \doi{10.1088/1361-6471/abf5c3}

\bibitem{Aguilar:2019teb}
A.C. Aguilar, et~al., Eur. Phys. J. A \textbf{55}(10), 190 (2019).
\newblock \doi{10.1140/epja/i2019-12885-0}

\bibitem{Conway:1989fs}
J.S. Conway, et~al., Phys. Rev. D \textbf{39}, 92 (1989).
\newblock \doi{10.1103/PhysRevD.39.92}

\bibitem{Aicher:2010cb}
M.~Aicher, A.~Schafer, W.~Vogelsang, Phys. Rev. Lett. \textbf{105}, 252003
  (2010).
\newblock \doi{10.1103/PhysRevLett.105.252003}

\bibitem{Cui:2020dlm}
Z.F. Cui, M.~Ding, F.~Gao, K.~Raya, D.~Binosi, L.~Chang, C.D. Roberts,
  J.~Rodr\'\i{}guez-Quintero, S.M. Schmidt, Eur. Phys. J. A \textbf{57}(1), 5
  (2021).
\newblock \doi{10.1140/epja/s10050-020-00318-2}

\bibitem{Cui:2020tdf}
Z.F. Cui, M.~Ding, F.~Gao, K.~Raya, D.~Binosi, L.~Chang, C.D. Roberts,
  J.~Rodr\'\i{}guez-Quintero, S.M. Schmidt, Eur. Phys. J. C \textbf{80}(11),
  1064 (2020).
\newblock \doi{10.1140/epjc/s10052-020-08578-4}

\bibitem{Saclay-CERN-CollegedeFrance-EcolePoly-Orsay:1980fhh}
J.~Badier, et~al., Phys. Lett. B \textbf{93}, 354 (1980).
\newblock \doi{10.1016/0370-2693(80)90530-4}

\bibitem{Cui:2021mom}
Z.F. Cui, M.~Ding, J.M. Morgado, K.~Raya, D.~Binosi, L.~Chang,
  J.~Papavassiliou, C.D. Roberts, J.~Rodr\'\i{}guez-Quintero, S.M. Schmidt,
  Eur. Phys. J. A \textbf{58}(1), 10 (2022).
\newblock \doi{10.1140/epja/s10050-021-00658-7}

\bibitem{Cui:2022bxn}
Z.F. Cui, M.~Ding, J.M. Morgado, K.~Raya, D.~Binosi, L.~Chang, F.~De~Soto, C.D.
  Roberts, J.~Rodr\'\i{}guez-Quintero, S.M. Schmidt, arXiv:2201.00884 [hep-ph] (2022)

\bibitem{Chang:2022jri}
L.~Chang, F.~Gao, C.D. Roberts, arXiv:2201.07870 [hep-ph] (2022)

\end{thebibliography}

\end{document}